\documentclass[fleqn,usenatbib]{mnras}

\usepackage[T1]{fontenc}
\usepackage{ae,aecompl}
\usepackage{float}

\usepackage{graphicx}	% Including figure files
\usepackage{amsmath}	% Advanced maths commands
\usepackage{amssymb}	% Extra maths symbols
\usepackage{float}
\usepackage{etoolbox}
\makeatletter
\patchcmd\@combinedblfloats{\box\@outputbox}{\unvbox\@outputbox}{}{\errmessage{\noexpand patch failed}}
\makeatother
\title[Four new ZZ Cetis]{Ground based observation of ZZ Ceti stars and the discovery of four new variables.}

\author[Romero et al.]{Alejandra D. Romero$^{1}$\thanks{E-mail: alejandra.romero@ufrgs.br},
L. Antunes Amaral$^{1}$,  T. Klippel$^{1}$,  D. Sanmartim$^{2}$, L Fraga$^{3}$, 
\newauthor G. Ourique$^{1}$,  I. Pelisoli$^{4}$,  G. R. Lauffer$^{1}$, S. O. Kepler$^{1}$ and D. Koester$^{5}$\\
$^{1}$Physics Institute, Universidade Federal do Rio Grande do Sul, Av. Bento Gon\c{c}alves 9500, Brazil\\
$^{2}$ Gemini Observatory, c/o AURA - Casilla 603, La Serena, Chile\\
$^{3}$ Laborat\'orio Nacional de Astrof\'isica LNA/MCTIC, 37504-364 Itajub\'a, MG, Brazil\\
$^{4}$ Institut f\"ur Physik und Astronomie, Universit\"atsstandort Golm, Karl-Liebknecht-Str. 24/25, 14467 Potsdam, Germany\\
$^{5}$ Institut f\"ur Theoretische Physik und Astrophysik, Universit\"at Kiel, D-24098 Kiel, Germany
}

\date{Accepted XXX. Received YYY; in original form ZZZ}

\pubyear{2018}

\begin{document}
\label{firstpage}
\pagerange{\pageref{firstpage}--\pageref{lastpage}}
\maketitle

% Abstract of the paper   200 words max.
\begin{abstract}
We perform ground based photometric observations of 22 DA white dwarf stars, 10 already known ZZ Cetis and 12 candidates with atmospheric parameters inside the classical instability strip. We report on the discovery of four new variable DA white dwarf stars. Two objects are near the middle of the instability strip, SDSS J082804.63$+$094956.6 and SDSS J094929.09$+$101918.8, and two red edge pulsators, GD 195 and L495$-$82.  In addition, we classified four objects as possible variables, since evidence of variability was detected in the light curve, but the S/N was not sufficient to establish a definite detection. Follow--up observations were performed for 10 know ZZ Ceti stars to verify period stability and search for new periodicities.  
For each confirmed variable, we perform a detailed asteroseismological fit and compare the structural parameters obtained from the best fit models with those obtained from spectroscopy and photometry from {\it Gaia}. Finally we present a study of the asteroseismological properties of a sample of 91 ZZ Ceti stars.

\end{abstract}


%%%%%%%%%%%%%%%%%%%%%%%%%%%%%%%%%%%%%%%%%%%%%%%%%%

%%%%%%%%%%%%%%%%% BODY OF PAPER %%%%%%%%%%%%%%%%%%

\section{Introduction}

ZZ Ceti stars are white dwarf stars with hydrogen dominated  atmospheres (DA) that show periodic variability. The instability strip of the ZZ Ceti is between 13 000 K and 10 000 K, depending on stellar mass \citep{2017ApJS..232...23H,2017EPJWC.15201011K}. Their photometric variations are due to surface temperature changes explained by spheroidal, non--radial g--modes pulsations with low harmonic degree ($\ell \leq 2$) and periods between 70 and 2000 s, with amplitude variations up to 0.3 mag. To date, there are $\sim 250$ ZZ Cetis known \citep[see][]{2016IBVS.6184....1B,2019arXiv190700115C}.

The driving mechanism for the excitation of the pulsations is the $\kappa-\gamma$ mechanism acting on the hydrogen partial ionization zone \citep{1981A&A...102..375D, 1982ApJ...252L..65W} for the blue edge of the instability strip. The convective driving mechanism \citep{1991MNRAS.251..673B, 1999ApJ...511..904G} is considered to be dominant once a thick convective zone has developed in the outer layers. 

The ZZ Cetis can be classified into three groups, depending on the effective temperature and the stellar mass \citep{2006ApJ...640..956M, 1993BaltA...2..407C}. The hot ZZ Cetis, which define the blue edge of the instability strip, exhibit a few modes with short periods ($< 350$ s) and small amplitudes (1.5--20 mma). The pulse shape is sinusoidal or sawtooth shaped and is stable for decades. On the opposite side of the instability strip are the cool DAV stars, showing several long periods (up to 1500 s), with large amplitudes (40--110 mma), and non sinusoidal light curves that change dramatically from season to season due to mode interference. \citet{2006ApJ...640..956M} suggested introducing a third class, the intermediate ZZ Cetis, with mixed characteristics from hot and cool ZZ Cetis.

Up until the Sloan Digital Sky Survey (SDSS), less than $\sim 50$ ZZ Cetis where known \citep[e.g.][]{2008PASP..120.1043F}, all with magnitudes $V< 16$. 
The number of known DA white dwarfs, and thus of DA pulsators, dramatically increased to $\sim 170$ members with the SDSS and the effort of several authors conducting ground based observations \citep{2004ApJ...607..982M, 2005ApJ...625..966M,2005A&A...442..629K,2012ApJ...757..177K,2006A&A...450..227C,2007A&A...462..989C, 2010MNRAS.405.2561C,2013MNRAS.430...50C,2013ApJ...779...58R}.

The list was enlarged with the discovery of pulsating white dwarfs stars within the {\it Kepler} spacecraft field\footnote{\url{http://archive.stsci.edu/kepler/data_search/search.php}}, thus opening a new avenue for white dwarf asteroseismology based on observations from space. This kind of data do not have the usual gaps due to daylight and also can cover months. However, the data reduction is quite challenging since a collection of instrumental frequencies, in the same range as those for known pulsators, need to be subtracted from the data \citep{2010ApJ...713L.160G,2013AcA....63..203B}. The first ZZ Ceti with published data was GD 1212 \citep{2014ApJ...789...85H}, already classified as variable by \citet{2006AJ....132..831G}, while the ZZ Ceti star observed the longest by the {\it Kepler} spacecraft was KIC 4552982, with data spanning more than 1.5 years.
%, discovered to be variable by ground based photometry by \citet{2011ApJ...741L..16H}. 
In particular, KIC 4552982 was the first ZZ Ceti to show energetic outbursts that increase the relative flux of the star  by 2\%-17\% \citep{2015ApJ...809...14B}. \citet{2017ApJS..232...23H} presented photometry and spectroscopy for 27 DAVs observed by the {\it Kepler} spacecraft, including six DAVs known at the time. They used this homogeneously analysed sample to study the white dwarfs rotation as a function of mass.

Data of similar quality to that provided by {\it Kepler} will be obtained by the \textit{Transiting Exoplanet Survey Satellite}\footnote{\url{https://heasarc.gsfc.nasa.gov/docs/tess/}} ({\it TESS}), launched in April 2018, which will perform a wide--field survey for planets that transit bright host stars \citep{2014SPIE.9143E..20R}. Compact pulsators, as white dwarfs and sub\-dwarf stars, will be studied with {\it TESS} since 2--minute cadence photometry is available \citep{2019RNAAS...3f..81B}. The acti\-vities related to compact pulsators are coordinated by the \textit{TESS} Compact Pulsators Working Group (WG\#8).
% Nearly $\sim$ 200 ?? candidates and known ZZ Ceti stars are part of the target list.
% The study of these object is coordinated by the TASC WG8 group (The TESS Asteroseismic Science Consortium Working Group 8). 

Time resolved ground based observations of variable white dwarf stars can help to increase the number of ZZ Ceti stars, and also other types of compact pulsators, to better understand the properties of ZZ Cetis and DA white dwarf stars in general. They can also function as a complement of space based surveys, given that in some cases, the resolution necessary to detect pulsations in variable DA white dwarfs is restricted to bright objects, especially for the ones near the blue edge of the instability strip. 
In addition, most of the known ZZ Ceti stars have pulsation periods only from the discovery observations. Follow--up observations of known pulsators can uncover new periodicities, improving the seismological studies. Finally, the stability of the pulsation modes, in amplitude and period, can carry information on the inner structure of the star as well \citep{2010ApJ...716...84M}.

In this paper, we carry out time series photometry observations of 22 DA white dwarfs. We performed follow--up observations on 10 known ZZ Ceti stars, and observe 12 ZZ Ceti candidates selected from spectroscopic parameters. For each object with confirmed variability, we perform a detailed asteroseismological fit by employing an expanded version of the grid of full evolutionary DA white dwarf models presented in \citet{2017ApJ...851...60R}.

This paper is organized as follows. We present our sample selection in Section \ref{sample}, describe the data reduction in Section \ref{observation} and present the observational results in Section \ref{obs-results}. In Section \ref{fits} we present our asteroseismological fits for the objects that show photometric variability. We present photometric determinations of effective temperature and stellar mass using {\it Gaia} magnitudes and parallax in Section \ref{gaia}. In Section \ref{astero-sample} we present a study of the asteroseismological properties of a sample of 77 ZZ Ceti stars, including the ones analysed in this work, that have been subject of an asteroseismological study. We conclude in section \ref{conclusion} by summarizing our findings. 
 
\section{Sample selection}
\label{sample}

\begin{figure*}
\includegraphics[width=0.85\textwidth]{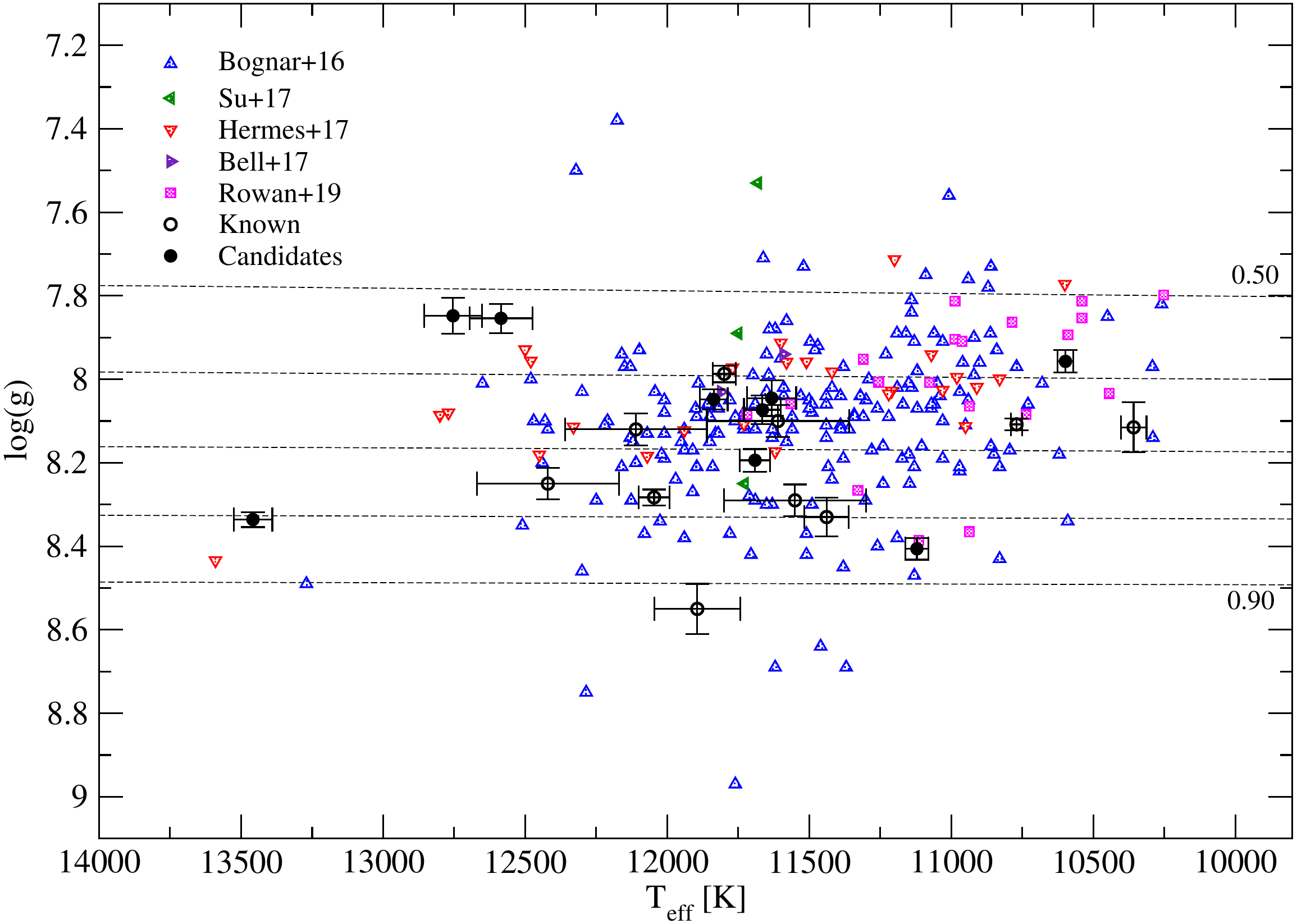}
\caption{Distribution of ZZ Ceti stars on the $T_{\rm eff}-\log g$ plane. Coloured symbols correspond to the ZZ Ceti stars known to date, extracted from \citet{2016IBVS.6184....1B} (blue triangle-up), \citet{2017ApJ...847...34S} (green triangle-left), \citet{2017ApJS..232...23H} (red triangle-down), \citet{2017ASPC..509..303B} (violet triangle-right) and \citet{2019MNRAS.486.4574R} (magenta square). The objects observed in this work are depicted with black circles, identified as candidates (full circle) and known variables (hollow circles). We include evolutionary tracks (dashed lines) with stellar masses between 0.5 and 0.9 $M_{\odot}$ from top to bottom \citep{2019MNRAS.484.2711R}. \label{Tlogg}}
\end{figure*}

We selected a list of targets from a sample of DA white dwarfs from the catalogues presented by \citet{2013ApJS..204....5K,2016MNRAS.455.3413K,2019MNRAS.486.2169K} from Sloan Digital Sky Survey  (SDSS). We choose those objects with spectroscopic effective temperature and surface gravity within the instability strip of the ZZ Cetis. In addition, we consider a sample of objects from the list of white dwarfs presented by \citet{2017ApJ...848...11B}, also with spectroscopic atmospheric parameters within the instability strip, that were not classified as variable white dwarfs. Finally, we selected a sample of known ZZ Ceti stars, most of them with published data corresponding only to the discovery paper, and no follow--up observations. A list of the objects observed in this work is presented in Table \ref{atmosphere}, where we list the spectroscopic effective temperature and surface gravity with and without 3D convection correction from \citet{2013A&A...552A..13T}. The stellar mass values were estimated by linear interpolation of the evolutionary tracks \citet{2012MNRAS.420.1462R,2013ApJ...779...58R} in the $\log g- T_{\rm eff}$ diagram, given the values of $\log g$ and effective temperature from table \ref{atmosphere}. We consider the spectroscopic values with 3D convection correction and the evolutionary sequences characterized by canonical hydrogen envelopes, i.e, those with the thickest value as predicted by single stellar evolution allowed by nuclear burning \citep[see][for details]{2019MNRAS.484.2711R}. The location of all observed objects on the $T_{\rm eff}-\log g$ plane is depicted in Figure \ref{Tlogg}. The $\sim$ 250 ZZ Cetis stars known to date are depicted in this figure, and were extracted from the works of \citet{2016IBVS.6184....1B} (blue up-triangle), \citet{2017ApJ...847...34S} (green left-triangle), \citet{2017ApJS..232...23H} (red down-tringle), \citet{2017ASPC..509..303B} (violet right-triangle) and \citet{2019MNRAS.486.4574R} (magenta square). The values for effective temperature and surface gravity were corrected by 3D convection for all objects \citep{2019arXiv190700115C}. The objects observed in this work are classified as candidates and known and depicted with full and hollow black circles, respectively.

\begin{table*}
	\centering
	\caption{Atmospheric parameters for the sample stars (columns 2 and 3), obtained from spectroscopy, and the stellar mass (column 4). The values corrected using the 3D
	convection correction \citep{2019arXiv190700115C} are listed in columns 5 and 6, and the resulting stellar mass is listed in column 7. Column 8 lists the references:
	(1) \citet{2011ApJ...743..138G}, (2) \citet{2013ApJS..204....5K}, (3) \citet{2017ApJ...848...11B} and (4) \citet{2019MNRAS.486.2169K}. }. 
	\label{atmosphere}
	\begin{tabular}{lccccccc} % four columns, alignment for each
		\hline
 star     &     $T_{\rm eff}$  &     $\log g$  &    $M\*/M_{\odot}$  &  $T_{\rm eff}$ 3D & $\log g$ 3D  &  $M\*/M_{\odot}$  3D &   ref.\\
\hline
BPM30551    & $11\,550 \pm 169$  &  $8.29 \pm  0.05$ & $0.7771 \pm 0.0339$  & $11\,240$ & 8.16 & $0.6936 \pm 0.0384$ & 1\\
%PG1149+058  & $11360 \pm 167$  &  $8.21 \pm  0.05$ & $0.7194 \pm 0.0318$  &&&& 1\\
HS1249+0426 & $12\,420 \pm 250$  &  $8.25 \pm  0.038$ & $0.7501 \pm 0.0385$  &$ 12\,160$ & 8.21 & $0.7204 \pm 0.0320$ & 1\\
HE1429$-$037  & $11\,610 \pm 178$  &  $8.10 \pm  0.05$ & $0.6597 \pm 0.0297$  & $11\,290$ & 8.00 & $0.6034\pm 0.0375$ & 1\\
GD385       & $12\,110 \pm 185$  &  $8.12 \pm  0.05$ & $0.6717 \pm 0.0303$  & $11\,820$ & 8.07 & $0.6429\pm 0.0388$ & 1\\
J2214$-$0025  & $11\,560 \pm  95$  &  $8.32 \pm  0.05$ & $0.7994 \pm 0.0382$  & $11\,650$ & 8.30 & $0.7826\pm 0.0446$ & 2\\
LP375-51      & $10\,076 \pm 148$  &  $8.000 \pm  0.050$ & $0.6005 \pm 0.0037$ &&&& 3\\
L495-82       & $11\,029 \pm 160$  &  $8.080 \pm 0.050$ & $0.6468 \pm 0.0393$ &&&& 3\\ 
%J0000$-$0046  & $11425 \pm  66$  &  $8.038 \pm  0.038$ & $0.6243 \pm 0.0293$ & 11472 & 8.181 & $0.6243 \pm 0.0293$ & 4\
SDSS J082804.63$+$094956.6  & $11\,673 \pm  53$  &  $8.067 \pm  0.027$ & $0.6409 \pm 0.0213$ & $11\,691$ & 8.194 & $0.6409 \pm 0.0213$ & 4\\     
SDSS J092511.63$+$050932.6  & $10\,874 \pm  20$  &  $8.332 \pm  0.014$ & $0.7999 \pm 0.0116$ & $10\,770$ & 8.108 & $0.6625 \pm 0.011$  & 4\\
SDSS J094929.09$+$101918.8  & $11\,685 \pm  65$  &  $8.202 \pm  0.034$ & $0.7161 \pm 0.0273$ & $11\,665$ & 8.073 & $0.6442 \pm 0.0268$ & 4\\
SDSS J095706.09$+$080504.8  & $12\,036 \pm  55$  &  $8.146 \pm  0.019$ & $0.6867 \pm 0.0159$ & $12\,046$ & 8.283 & $0.7740 \pm 0.0169$ & 4\\
SDSS J113325.69$+$183934.7  & $11\,223 \pm  40$  &  $8.603 \pm  0.026$ & $0.9695 \pm 0.0221$ & $11\,121$ & 8.406 & $0.8465 \pm 0.023$  & 4\\
WD1345$-$0055 & $11\,799 \pm  40$  &  $8.095 \pm  0.020$ & $0.6572 \pm 0.0159$ & $11\,799$ & 7.987 & $0.5976 \pm 0.0151$ & 4\\
WD1451$-$0111 & $13\,369 \pm  68$  &  $8.362 \pm  0.018$ & $0.8213 \pm 0.0160$ & $13\,458$ & 8.336 & $0.8055 \pm 0.0151$ & 4\\         
GD 195        & $11\,833 \pm  49$  &  $8.163 \pm  0.024$ & $0.6967 \pm 0.0194$ & $11\,836$ & 8.048 & $0.6309 \pm 0.0185$ & 4\\
SDSS J161005.17$+$030256.1  & $12\,649 \pm  92$  &  $7.877 \pm  0.043$ & $0.5429 \pm 0.0284$ & $12\,754$ & 7.848 & $0.5296 \pm 0.022 $ & 4\\  
SDSS J161218.08$+$083028.1  & $12\,668 \pm 76$   &  $8.549 \pm 0.021$ &  $0.9373 \pm 0.0093$ & $12\,722$ & 8.441 & $0.8709 \pm 0.0094$ & 4\\
SDSS J212441.27$-$073234.9 &  $13\,991\pm 164$ & $7.834\pm 0.035$ & $0.5265\pm 0.0163$ & $14\,069$ & 7.835 & $0.5271\pm 0.0163$ & 4 \\  
SDSS J213159.88$+$010856.3  & $11\,655 \pm  86$  &  $8.172 \pm  0.043$ & $0.7020 \pm 0.0329$ & $11\,632$ & 8.045 & $0.6387 \pm 0.0332$ & 4\\
SDSS J215905.53$+$132255.8  & $11\,941 \pm 151$  &  $8.713 \pm  0.060$ & $1.0249 \pm 0.0390$ & $11\,894$ & 8.550 & $0.9385 \pm 0.0525$ & 4\\
SDSS J235040.72$-$005430.9  & $10\,468 \pm  45$  &  $8.380 \pm  0.060$ & $0.8292 \pm 0.0531$ & $10\,358$ & 8.115 & $0.6656 \pm 0.0493$ & 4\\
SDSS J235932.80$-$033541.1  & $10\,706 \pm  28$  &  $8.196 \pm  0.027$ & $0.7122 \pm 0.0200$ & $10\,598$ & 7.957 & $0.5788 \pm 0.020$  & 4\\
\hline
	\end{tabular}
\end{table*}

\section{Observations and Data Reduction}
\label{observation}

%We carried out time series photometry using two different telescopes during the years of 2015 to 2019.
We employed Goodman image mode on the 4.1-m Southern Astrophysical Research (SOAR) Telescope from 2015 to 2019. We used read out mode 200 Hz ATTN2 with the CCD binned 2$\times$2. All observations were obtained with a red blocking filter S8612. The integration times varies from 10 to 60 sec, depending on the magnitude of the object and the weather conditions. 

In addition, we used the IxON camera on the 1.6-m Perkin Elmer Telescope at the Pico dos Dias Observatory during 2016, 2017 and 2018. We also used a red blocking filter BG40. The integration times varies from 20 to 45 sec, depending on the magnitude of the object. The journal of observations is shown in Table \ref{journal}.

\begin{table*}
	\centering
	\caption{Journal of observations for the objects observed. $\Delta t$ is the total length of each observing run and $t_{\rm exp}$ is the integration time of each exposure.}
	\label{journal}
	\begin{tabular}{lccccccc} % four columns, alignment for each
		\hline
 star     &  RA  & DEC  & $g$ & Telescope  &    Run start (UT)  &  $t_{\rm exp}$ (sec) &  $\Delta t$ (h) \\
\hline
Known variables             &                 &   &             &  &                 &    & \\
\hline
%J0000$-$0046 & 00 00 06.75 & -00 46 53.98 & 18.66 & SOAR & 2016-08-22 06:33:07.32 & 20 & 2.00 \\
BPM 30551    & 01 06 53.68  & -46 08 53.73 & 15.47 & SOAR & 2016-08-23 08:35:52.83 & 10 & 1.55 \\
             &              &              &       & OPD  & 2016-08-30 06:20:21.12 & 10 & 2.08 \\
SDSS J092511.63$+$050932.6 & 09 25 11.63  & +05 09 32.6  & 15.20 & OPD  & 2016-04-17 22:41:16.91 & 35 & 2.47 \\
             &              &              &       & OPD  & 2016-04-17 00:19:40.72 & 35 & 1.00 \\
HS1249$+$0426 & 12 52 15.19 & +04 10 52.9 & 16.04 & OPD  & 2016-04-16 03:46:00.40 & 30 & 1.90 \\
              &             &             &       & OPD     & 2016-04-18 03:27:02.32 & 45 & 2.03 \\
WD1345$-$0055 & 13 45 50.92 & -00 55 36.4 & 16.78 & OPD  & 2016-04-17 03:56:59.02 & 15 & 2.50 \\
HE1429$-$037 & 14 32 03.19  & -03 56 38.2 & 16.03 & OPD  & 2017-04-17 03:28:43.74 & 30 & 1.87 \\
SDSS J161218.08$+$083028.1 & 16 12 18.08  & +08 30 28.1  & 17.75 & SOAR & 2014-07-02 00:11:16.35 & 30 & 1.87 \\
GD 385       & 19 52 27.88  & +25 09 29.10 & 16.63 & OPD  & 2016-08-29 23:08:02.08 & 10 & 3.47 \\
SDSS J215905.53$+$132255.8 & 21 59 05.53  & +13 22 55.8  & 18.99 & SOAR & 2016-08-22 01:38:56.16 & 30 & 2.00 \\
SDSS J221458.37$-$002511.9 & 22 14 58.37  & -00 25 11.91 & 17.92 & OPD  & 2016-08-29 01:58:31.27 & 30 & 4.12 \\
SDSS J235040.72$-$005430.9 & 23 50 40.72  & -00 54 30.87 & 18.12 & SOAR & 2016-08-22 04:17:18.89 & 15 & 2.00 \\
\hline
Variable candidates            &              &  &             &  &              &    & \\
\hline
SDSS J082804.63$+$049456.6 & 08 28 04.63  & +09 49 56.66 & 17.71 & SOAR & 2016-12-24 03:57:54.27 & 15 & 2.06 \\ 
             &              &              &       & SOAR & 2016-12-27 04:16:49.31 & 15 & 4.19 \\
SDSS J094929.09$+$101918.85 & 09 49 29.09  & +10 19 18.85  & 17.58 & SOAR & 2016-12-24 06:09:23.07 & 15 & 2.07 \\
             &              &              &       & SOAR & 2017-01-29 07:16:50.90 & 15 & 1.29 \\
             &              &              &       & SOAR  & 2015-03-19 00:15:51.21  & 30  & 4.26 \\
SDSS J095703.09$+$080504.8 & 09 57 03.09  & +08 05 04.85 & 17.70 & OPD  & 2017-04-15 01:58:40.26 & 40 & 1.16 \\
             &              &              &       & OPD  & 2017-04-16 22:09:13.03 & 40 & 4.50 \\
             &              &              &       & SOAR & 2017-01-29 03:45:34.29 & 20 & 3.11 \\  
SDSS J113325.09$+$183934.7 & 11 33 25.69  & +18 39 34.75 & 17.59 & OPD  & 2017-04-15 00:37:22.27 & 50 & 2.65 \\
             &              &              &       & OPD  & 2018-05-10 22:00:03.75 & 20 & 4.14 \\
LP 375$-$51  & 11 50 20.17  & +25 18 32.76 & 15.70 & OPD  & 2018-05-11 23:44:21.80 & 30 & 2.66 \\  
WD1454$-$0111 & 14 54 36.08 & -01 11 52.5  & 17.34 & OPD  & 2016-04-15 06:14:45.5  & 30 & 2.22 \\
GD 195       & 16 07 46.21  & +17 37 20.76 & 16.63 & OPD  & 2016-04-18 06:09:35.46 & 50 & 2.33 \\
             &              &              &       & OPD  & 2016-04-18 05:51:50.89 & 40 & 2.65 \\ 
             &              &              &       & OPD  & 2016-04-17 07:04:47.12 & 40 & 1.44 \\
             &              &              &       & OPD  & 2017-04-16 05:16:58.56 & 20 & 3.00 \\ 
SDSS J161005.17$+$030256.1 & 16 10 05.17  & +03 02 56.07 & 18.55 & SOAR & 2017-08-06 23:08:09.69 & 15 & 3.00 \\ 
L495$-$82    & 20 43 49.2   & -39 03 18.2  & 13.76 & OPD  & 2018-05-12 05:11:34.94 & 10 & 2.85 \\
SDSS J212441.27$-$073234.9 & 21 24 41.27  & -07 32 34.93 & 18.47 & SOAR & 2019-05-21 07:31:23.14 & 60 & 2.44 \\
             &              &              &       & SOAR & 2019-05-22 07:32:59.99 & 22 & 2.60 \\   
SDSS J213159.88$+$010856.3 & 23 31 59.88  & +01 08 56.26 & 18.40 & SOAR & 2017-08-07 02:44:00.29 & 30 & 1.88 \\
             &              &              &       & SOAR & 2017-07-13 07:08:01.08 & 30 & 2.40 \\
SDSS J235932.80$-$033541.1 & 23 59 32.80  & -03 35 41.07 & 17.91 & SOAR & 2017-07-22 09:02:34.37 & 10 & 1.58 \\
 
\hline 
            
	\end{tabular}
\end{table*}

%\section{Data reduction and light curves}

We reduced the data with the software {\tt IRAF}, and perform aperture photometry with {\tt DAOFOT}. We extracted light curves of all bright stars that were observed simultaneously in the field. Then, we divided the light curve of the target star by the light curves of all comparison stars to minimize effects of sky and transparency fluctuations. To look for periodicities in the light curves, we calculate the Fourier Transform (FT) using the software {\tt Period04} \citep{2004IAUS..224..786L}. We accepted a frequency peak as significant if its amplitude exceeds an adopted significance threshold. In this work, we adopted a  4<A> significance criterion, where <A> is the mean amplitude of the FT, corresponding to a probability of the peak being due to noise smaller than 1 in 1000 \citep{1993BaltA...2..515K}. We then use the process of pre--whitening the light curve by subtracting out of the data  a sinusoid with the same frequency, amplitude and phase of highest peak and then computing the FT the residuals. We redo this process until we have no new significant signals. The objects classified as candidates were observed for a minimum of three hours in total to confirm variation. As a result, we find four new ZZ Ceti stars among the candidates. We also discovered new periods for some known variables.

\section{Observational results}
\label{obs-results}

In this section, we present the results from the observations for the 22 objects observed for this work. We found four new ZZ Ceti stars and four possible new variables. For the known pulsators, we recovered most of the periods from the literature and detected new modes. From the FT we were not able to detect any multiplets to extract information on the harmonic degree. Finally, in the case of the rich pulsators, we looked for linear combinations among the detected periodicities, to select those periods corresponding to real pulsation modes. We detail the results from the observations below.

\subsection{New ZZ Cetis}

From the observed sample, we found four new ZZ Ceti stars: SDSS J082804.63$+$094956.6, SDSS J094929.09$+$101918.8 , GD 195 and L495$-$82. We present the results for each object below. The light curves and Fourier Transform for each object are depicted in Figures \ref{FT1}, \ref{FT2}, \ref{FT3} and \ref{FT4}, respectively, while the list of observed frequencies, periods and amplitudes is presented in Table \ref{new-periods}.

\subsubsection{SDSS J082804.63$+$094956.6}

\begin{figure}
\includegraphics[width=\columnwidth]{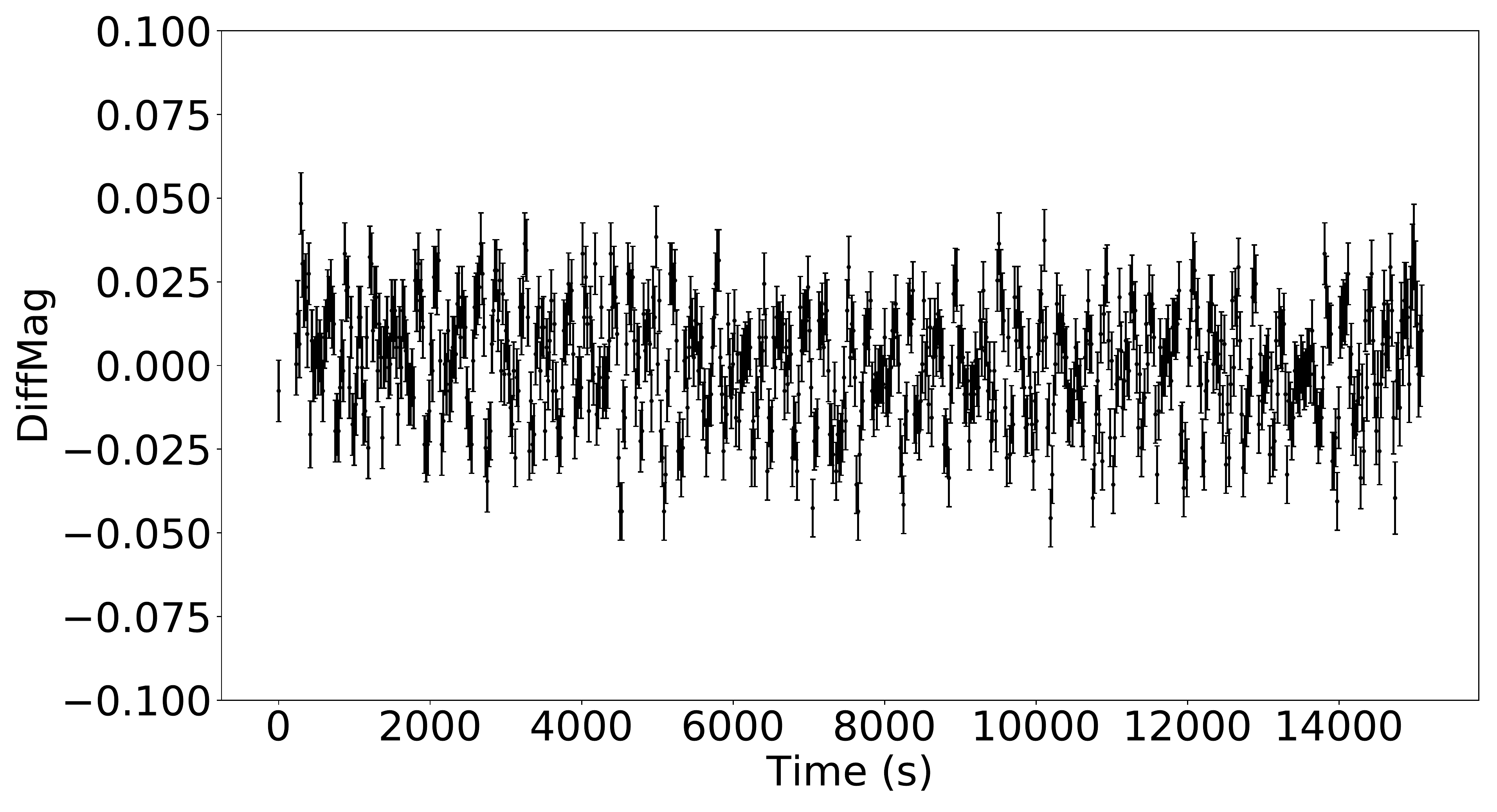}
\includegraphics[width=\columnwidth]{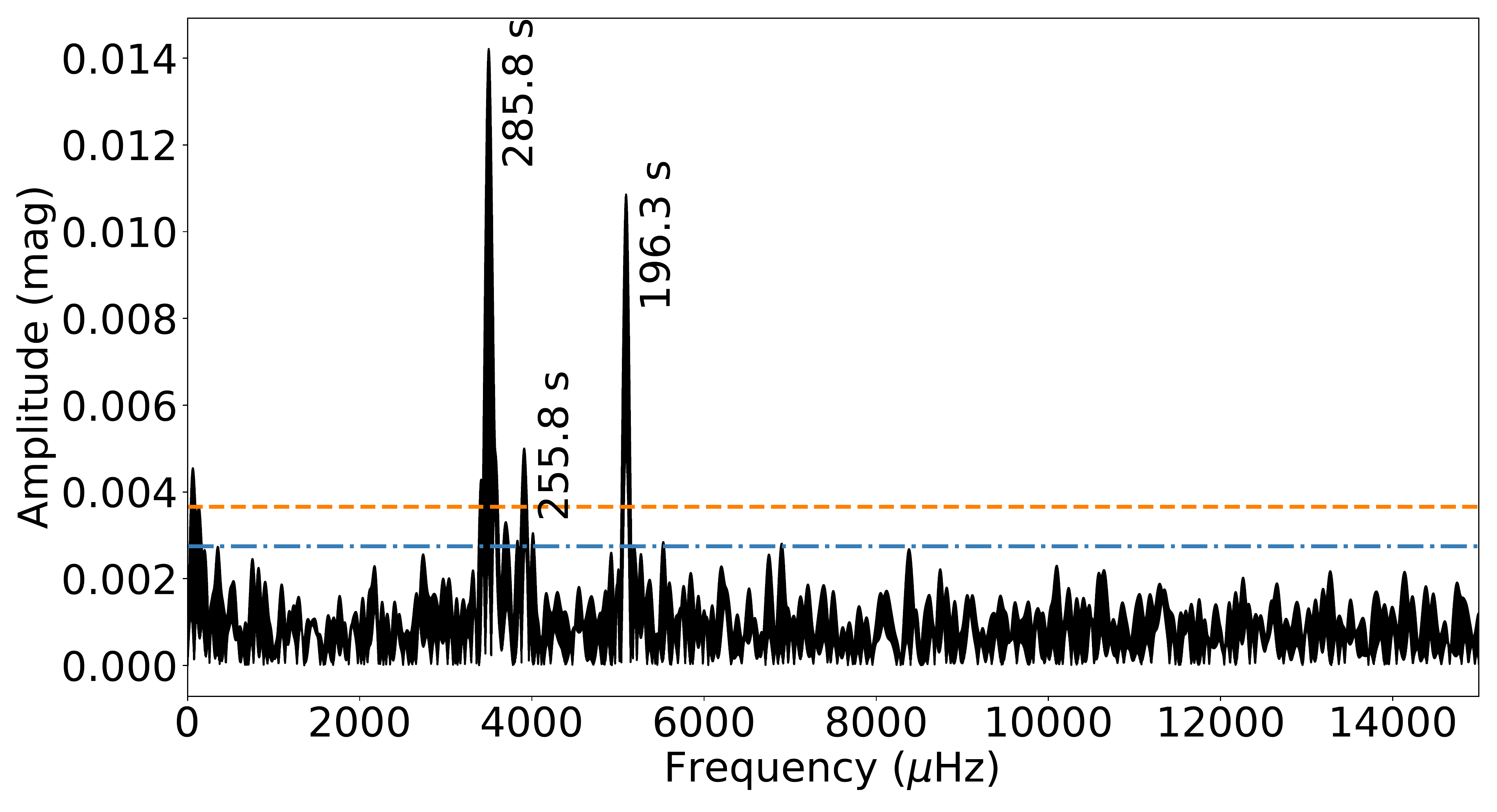}
\caption{ Light curve (top panel) and Fourier Transform (bottom panel) for the object SDSS J082804.63$+$094956.6. The light curve corresponds to the 4 h run, while for the FT we consider the two observation nights. The orange dashed (blue dotted-dashed) line correspond to the 4$\sigma$ (3$\sigma$) detection limit.  \label{FT1}}
\end{figure}

The star J0828$+$0949 was selected as a candidate from the SDSS catalogue presented by  \citet{2016MNRAS.455.3413K}. It was observed in two nights for a total of 6 h with the SOAR telescope. In Figure \ref{FT1} we show the light curve for the 4 h run (top panel), and the Fourier Transform corresponding to all observation nights (bottom panel), where the dashed and dotted-dashed lines correspond to the 3$\sigma$ and 4$\sigma$ limit, respectively.  The FT shows three well defined peaks above the 4$\sigma$ limit in the high frequency domain. The detected frequencies and periods and the corresponding amplitudes are listed in Table \ref{new-periods}. The modes show short periods between 196 and 285 sec, corresponding to a blue edge pulsator. From the FT we were not able to detect any multiplets, so no harmonic degree can be obtained directly from the observations.

\subsubsection{SDSS J094929.09+101918.8}

\begin{figure}
\includegraphics[width=\columnwidth]{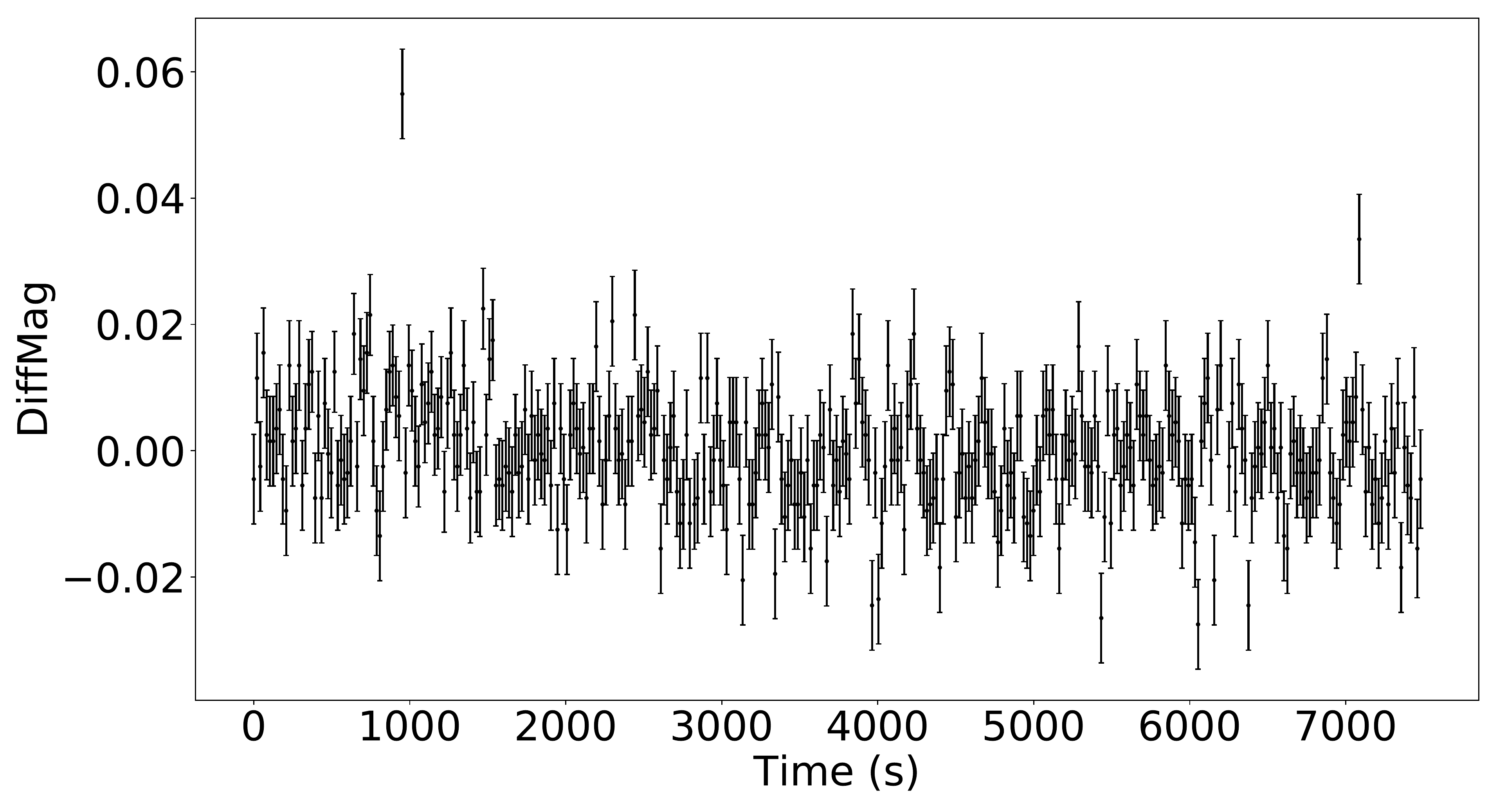}
\includegraphics[width=\columnwidth]{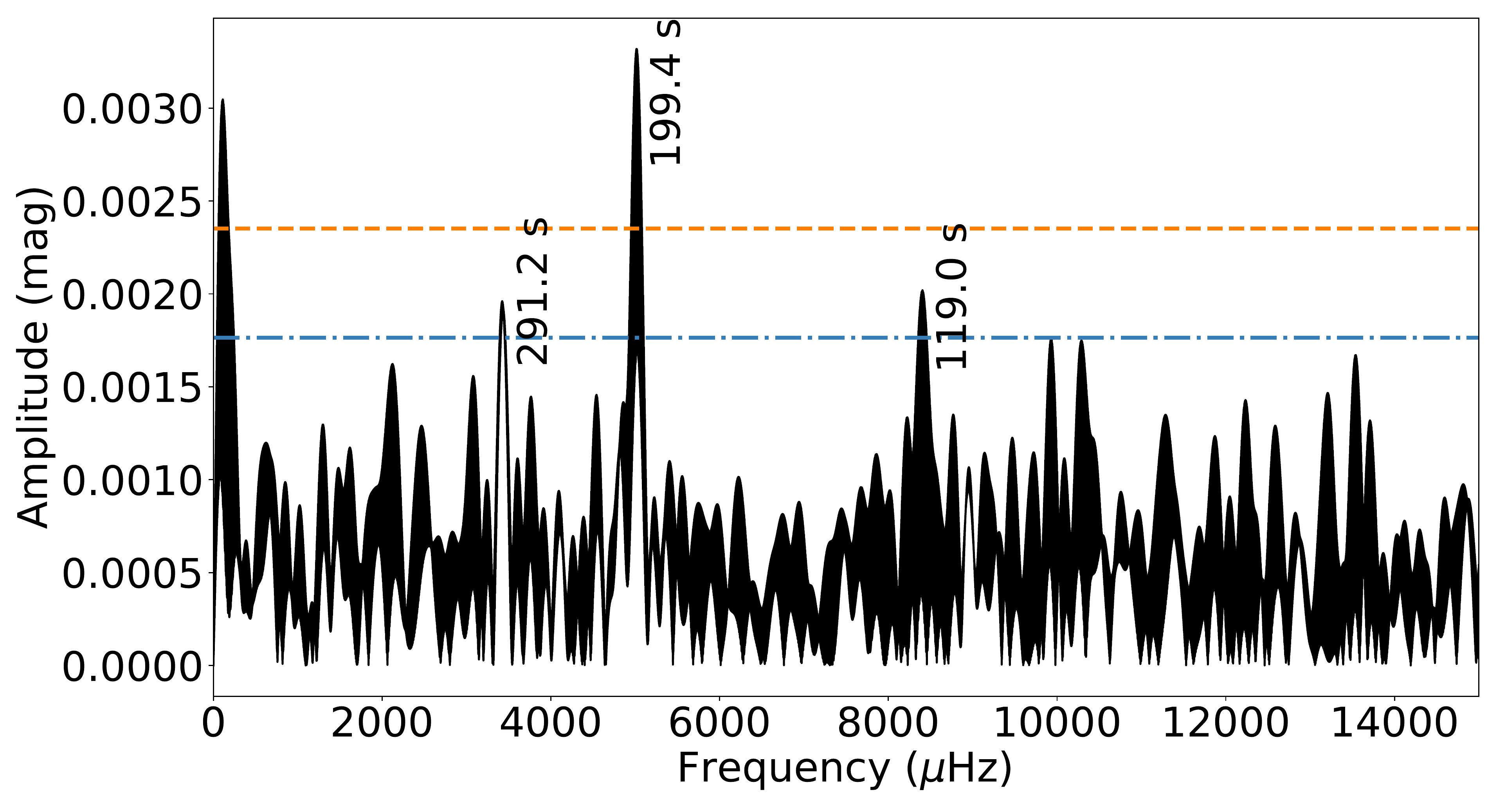}
\caption{Light curve (top panel) and Fourier Transform (bottom panel) for the star SDSS J094929.09+101918.8. The light curve corresponds to the 2.07 h observation run. The Fourier Transform is the result from the sum of both nights. The orange dashed (blue dotted-dashed) line correspond to the 4$\sigma$ (3$\sigma$) detection limit. \label{FT2}}
\end{figure}

SDSS J094929.09+101918.8 was also selected from the SDSS catalogue presented by  \citet{2016MNRAS.455.3413K,2019MNRAS.486.2169K}. From spectroscopy the star has a stellar mass of $0.644 M_{\odot}$ and a 3D effective temperature of $11\, 665$ K. It was observed for a total of 3.36 h on the SOAR telescope. In  figure \ref{FT2} we present the light curve for the 2.07 h observation run and the FT for all observations. As can be seen from the FT, this object shows one period of 199.31 s above the 4$\sigma$ limit. For amplitudes lower than 4$\sigma$, but higher than 3$\sigma$, we found two additional periods. In particular, the mode with a period of 291.20 s is present only in the second observation night, while the period of 119 s appears when we combine all observations, and it is a linear combination of the other two modes. 

\subsubsection{GD 195}

\begin{figure}
\includegraphics[width=\columnwidth]{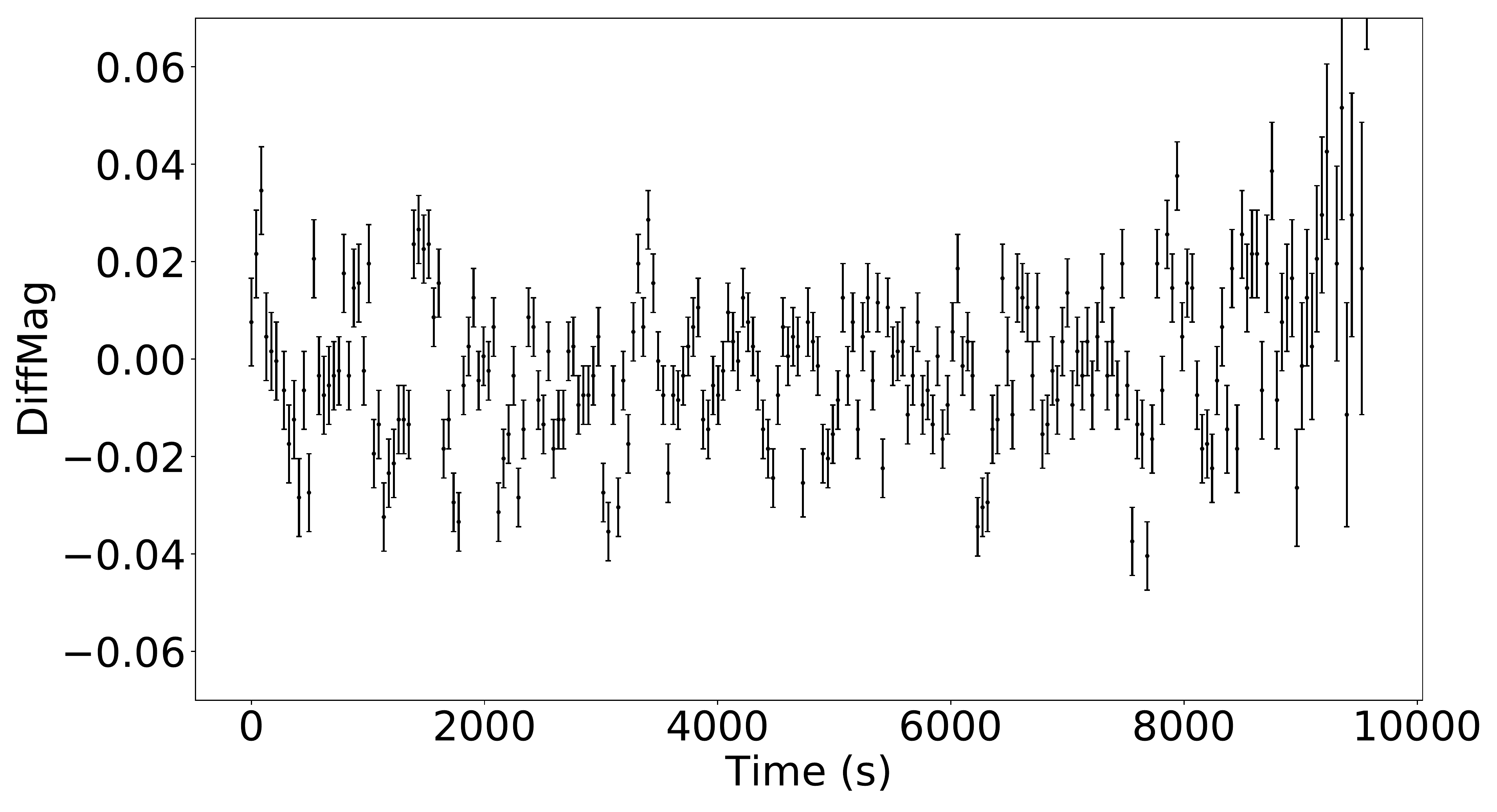}
\includegraphics[width=\columnwidth]{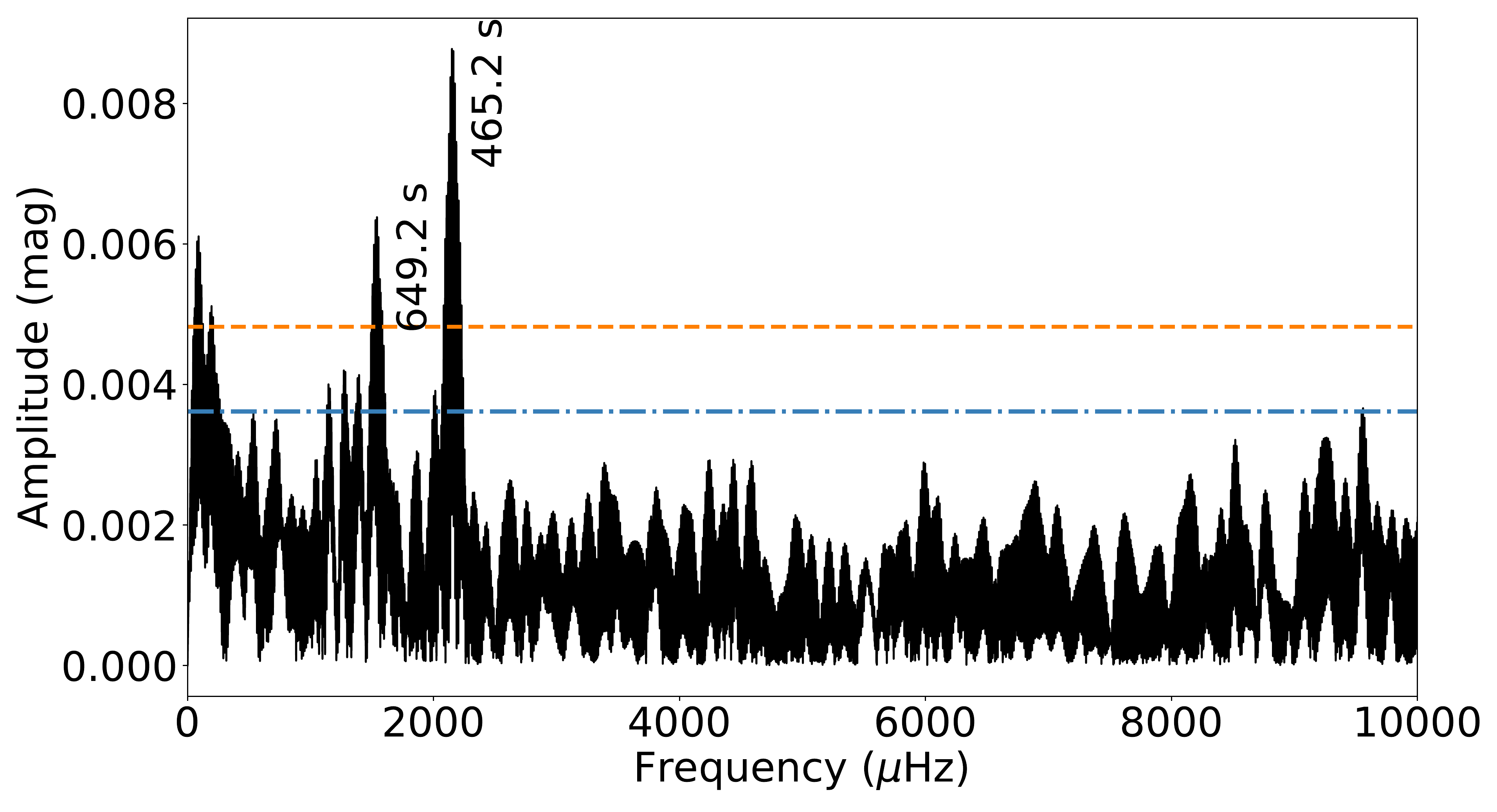}
\caption{Light curve (top panel) and  Fourier Transform (bottom panel) for GD 195. The light curve corresponds to the 2.65 h run, while the Fourier Transform corresponds to the sum of all observations. The orange dashed (blue dotted-dashed) line correspond to the 4$\sigma$ (3$\sigma$) detection limit. \label{FT3}}
\end{figure}

GD 195 was classified as a very hot white dwarf star by \citet{2011ApJ...743..138G}, with atmospheric parameters of $T_{\rm eff} = 14\, 590\pm277$ K and $\log g = 7.82\pm 0.05$. However, \citet{2016MNRAS.455.3413K} found an effective temperature of $T_{\rm eff} = 11\, 833 \pm 49$ K and $\log g = 8.163\pm0.024$ based on SDSS spectra fitted with an updated version of the atmospheric models from \citet{2010MmSAI..81..921K} \citep[see also][]{2019MNRAS.486.2169K}. 
Additional fitting using a grid of updated models from \citet{2010MmSAI..81..921K} with $\alpha = 0.8$ and 0.7, showed that the spectroscopic effective temperature of GD 195 is lower than $12\, 000$ K, putting the star inside the classical ZZ Ceti instability strip \citep[see][for details on the fitting procedure]{2019MNRAS.486.2169K}.
%Additional fitting using updated models from \citet{2010MmSAI..81..921K} with different mixing length $\alpha$ showed that the effective temperature of GD 195 is lower than $12\, 000$ K, putting the star inside the classical ZZ Ceti instability strip. 
Figure \ref{FT3} shows the FT corresponding to the combination of the three nights of observations. We found two modes with periods of 465.16 and 649.20 s. According to the classification presented by \citet{2006ApJ...640..956M}, a ZZ Ceti with periods longer than $\sim 350$ s correspond to a variable in the middle of the instability strip with $T_{\rm eff}\sim 11600$ K. This is in better agreement with the spectroscopic determinations of \citet{2016MNRAS.455.3413K}.

%Finally, using GAIA  observations of magnitude and parallax, combined with models from Bergeron, we find a stellar mass of ~0.4-0.5 $M_{\odot}$ and an effective temperature of $\sim 10700$.

\subsubsection{L495$-$82}

\begin{figure}
\includegraphics[width=\columnwidth]{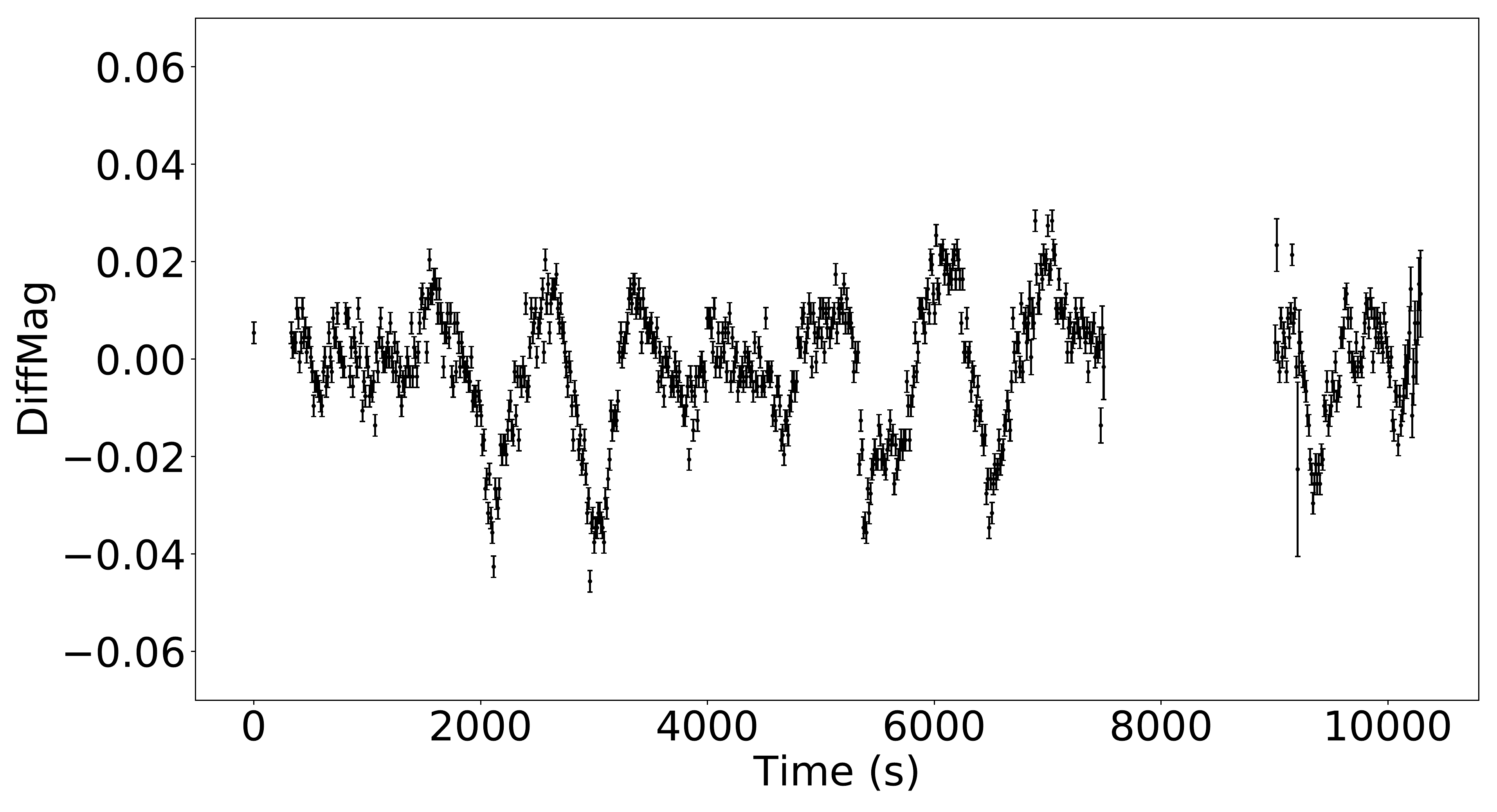}
\includegraphics[width=\columnwidth]{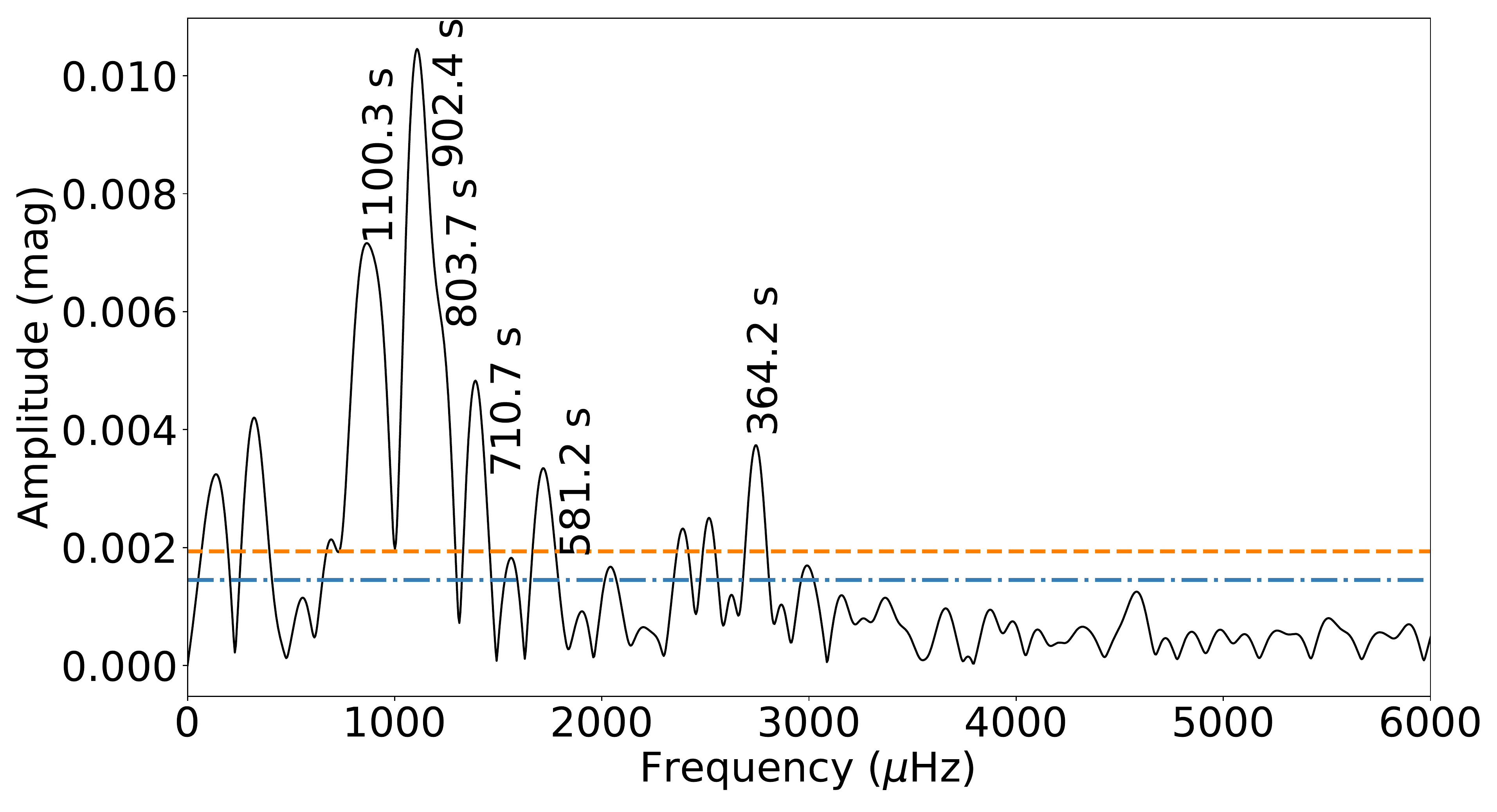}
\caption{Light curve (top panel) and Fourier Transform (bottom panel) for the star L495$-$82. The orange dashed (blue dotted-dashed) line correspond to the 4$\sigma$ (3$\sigma$) detection limit.  \label{FT4}}
\end{figure}

L495$-$82 was selected from the list of objects presented by \citet{2017ApJ...848...11B}. This star is quite bright as compared to the other observed targets, with a $g$ magnitude of 13.76. Figure \ref{FT4} depicts the light curve and Fourier Transform for L495$-$82, that shows a collection of long period modes, with a dominant period in 902.4 s, compatible with a red edge pulsator \citep{1993BaltA...2..515K}. As can be seen from Table \ref{new-periods} there are several linear combinations which is also characteristic of a cool ZZ Ceti.

Since we observed this object for only $\sim$3 h, we cannot define all modes accurately, especially around the dominant peak corresponding to a period of 902 s. In particular, the mode f$_3$ is close to the linear combination (f$_1+$f$_4$)/2 = 1256.36 $\mu$Hz, which is within the uncertainties given that the peak for f$_3$ has a width of 30$\mu$Hz. Thus, for asteroseismological purposes we will consider the f$_4$ as a real mode and f$_3$ as a linear combination.

\begin{table}
	\centering
	\caption{Detected modes for the new pulsators. Column 1 lists the name, while the frequency, period and amplitude are listed in columns 2, 3 and 4, respectively. Column 5 shows the identification of the mode, if it is a normal mode or a linear combination.}
	\label{new-periods}
	\scalebox{0.85}[0.85]{ \hspace{-7mm}
	\begin{tabular}{lcccc} % four columns, alignment for each
		\hline
 star     &   Freq   & Period  &  Amp & ID\\
          &  ($\mu$Hz) & (s) & (mma) & \\
\hline
SDSS J082804.63$+$094956.6  &  3499.073 &  285.79  &  14.0 & f$_1$ \\
              &  5093.984 &  196.31  &  10.9 & f$_2$ \\
              &  3909.610 &  255.78  &   5.0 & f$_3$ \\           
\hline
SDSS J094929.09$+$101918.8  &  5017.309 &  199.31  &   4.7 & f$_1$ \\
              &  3434.066 &  291.20  &   1.9 & {\it f$_2$}\\  
              &  8403.361 &  119.00  &   2.0 & {\it f$_1+$f$_2$}\\
\hline
GD 195      &  2149.798 &  465.16  &  8.7 & f$_1$ \\
            &  1540.357 &  649.20  &  6.2 & f$_2$\\
\hline
L495$-$82   &  1108.125 &  902.425 & 10.48 & f$_1$ \\ 
            &   908.856 & 1100.283 &  6.72 & f$_2$  \\ 
            &  1244.210 &  803.722 &  5.74 & f$_3$\\
            &  1404.597 &  711.947 &  3.92 & f$_4$ \\
            &  2746.011 &  364.164 &  3.59 & f$_5$ \\
            &   991.480 & 1008.590 &  3.74 & f$_6$ \\
            &  1720.509 &  581.223 &  2.62 & f$_7$ \\
            &  2396.375 &  417.296 &  2.00 & f$_4+$f$_6$ \\
            &  4578.305 &  218.421 &  1.41 & {\it f$_8$}  \\
            &  2031.562 &  492.23  &  1.39 & {\it f$_9$} \\
            &  3018.182 &  331.32  &  1.21 & f$_6+$f$_{10}$ \\
            &  2541.883 &  393.409 &  1.19 & f$_8-$f$_9$ \\
\hline             
	\end{tabular}}
\end{table}

\subsection{Known pulsators}

\begin{table*}
	\centering
	\caption{List of periods characteristic of the known variables observed in this work. We list the periods, in sec, detected in this work and their amplitudes (in mma) is columns 2 and 3, while the periods detected in previous works and their amplitudes are listed in columns 4 and 5, along with the references in column 6.}
	\label{known-old}
	\begin{tabular}{lccccc} % four columns, alignment for each
		\hline
 star     &  This work &&  known & & \\
\hline
        &  Period  & amp & Period  & amp  & Ref. \\
\hline
BPM 30551    &  831.031 & 11.2 & 606.8 & 11.5 & \citet{1976ApJ...209..853H} \\
             &  775.235 & 11.3 & 744.7 & 10.5 &  \\
             &  959.780 &  7.7 & 682.7 & $\sim$10 &\\
             &  460.060 &  5.7 & 840.2 & $\sim$10 &\\
             &  986.357 &  5.5 &       &          & \\ 
             &  649.348 &  5.5 &       &          & \\     
\hline
SDSS J092511.63$+$050932.6 & 1247.46 &  8.0 & 1127.14 & 3.17 & \citet{2010MNRAS.405.2561C} \\
             &         &      & 1264.29 & 3.05 & \\
             &         &      & 1159.00 & 2.7 & \citet{2013ApJ...779...58R} \\
             &         &      & 1341.00 & 4.0 & \\
\hline
HS 1249$+$0426 & 294.91 & 14.5 & 288.9 & 7.6 & \citet{2006AA...450.1061V} \\
\hline
WD1345$-$0055 & 195.24 & 8.9 & 195.2  & 5.5 & \citet{2004ApJ...607..982M} \\
              &        &     & 254.4  & 2.4 & \\       
\hline
HE 1429$-$037 & 821.74  & 56.93 & 450.1  & 10.2 & \citet{2005AA...443..195S} \\
             &         &       & 826.4  & 18.3 & \\
             &         &       & 969.0  & 12.7 & \\
             &         &       & 1084.9 & 16.3 & \\  
\hline
SDSS J161218.08$+$083028.1 & 115.122 & 5.14 & 115.17 & 5.06 &  \citet{2013MNRAS.430...50C}\\
\hline
GD 385       & 256.09  & 9.4 & 256.12 & 11.4 & \citet{2009MNRAS.396.1709C} \\
             & 127.93 & 3.5  & 128.15 & 3.7 & \\
\hline
SDSS J215905.53$+$132255.8 & 678.78 & 8.0 & 683.7 & 11.7 & \citet{2005ApJ...625..966M} \\
             & 746.67 & 24.2 & 801.0 & 15.1 & \\
\hline
SDSS J221458.37$-$002511.9 & 255.08 & 16.0 & 255.2 & 13.1 & \citet{2005ApJ...625..966M} \\
             &        &      & 195.2 & 6.1 & \\ 
\hline
SDSS J235040.72$-$005430.9 & 304.74 & 18.29 & 304.3 & 17.0 & \citet{2004ApJ...607..982M} \\
             & 390.32 & 10.17 & 391.1 & 7.5 &\\
             & 271.87 & 8.2   & 273.3 & 6.2 & \\
             &        &       & 206.7 & 3.2 & \citet{2006ApJ...640..956M}\\ 
\hline            
	\end{tabular}
\end{table*}

\begin{figure}
\includegraphics[width=\columnwidth]{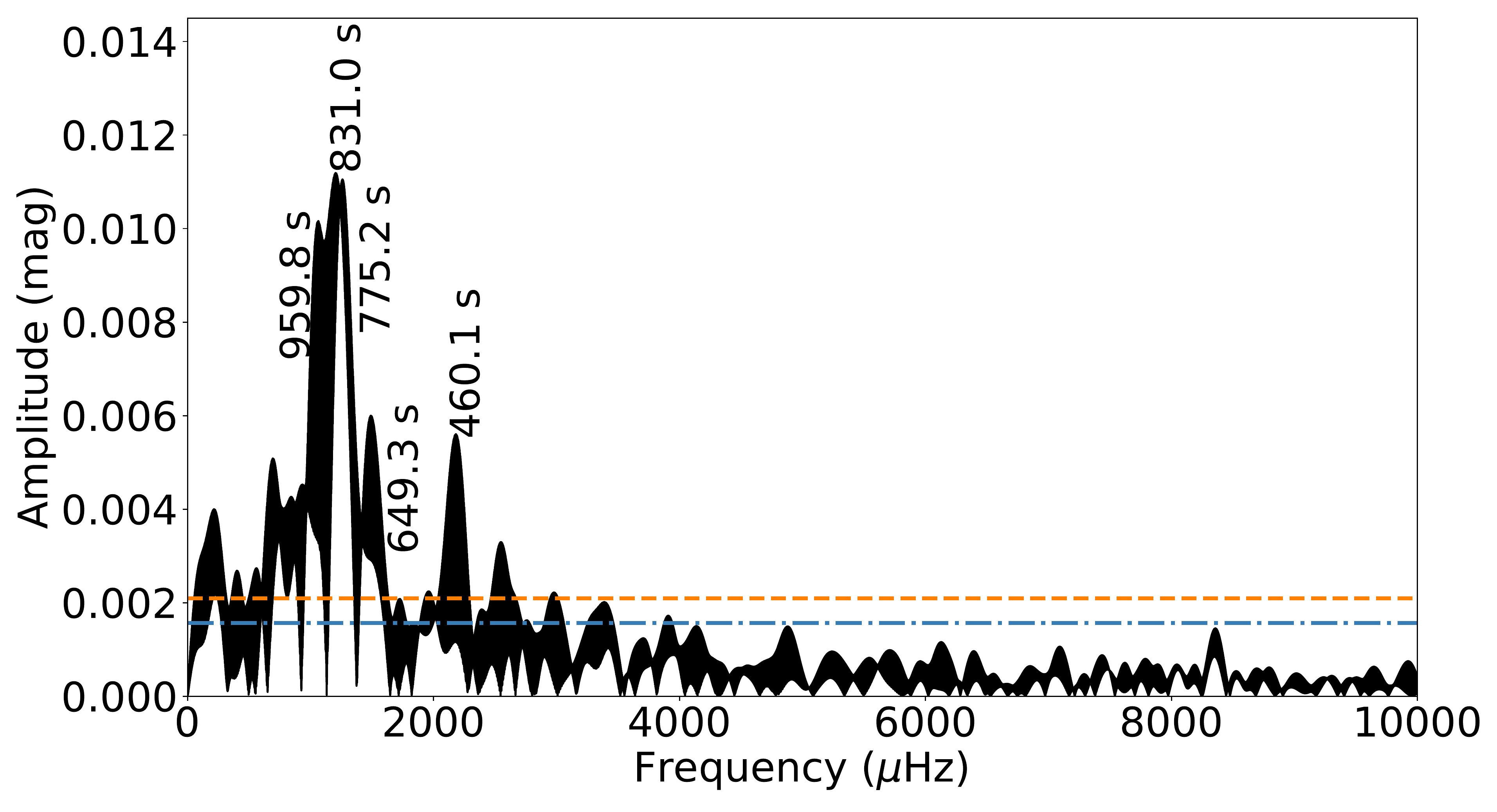}
\includegraphics[width=\columnwidth]{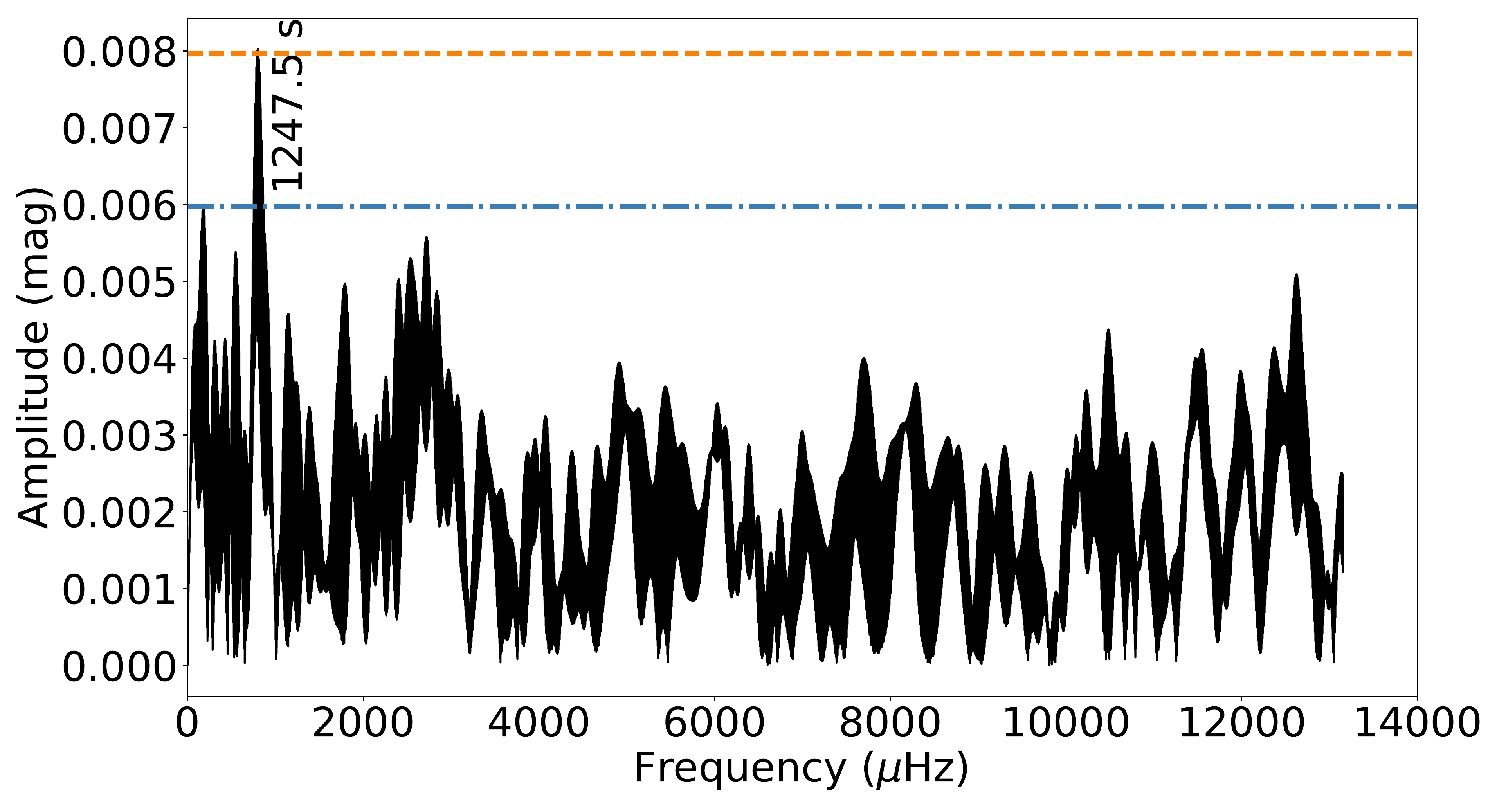}
\includegraphics[width=\columnwidth]{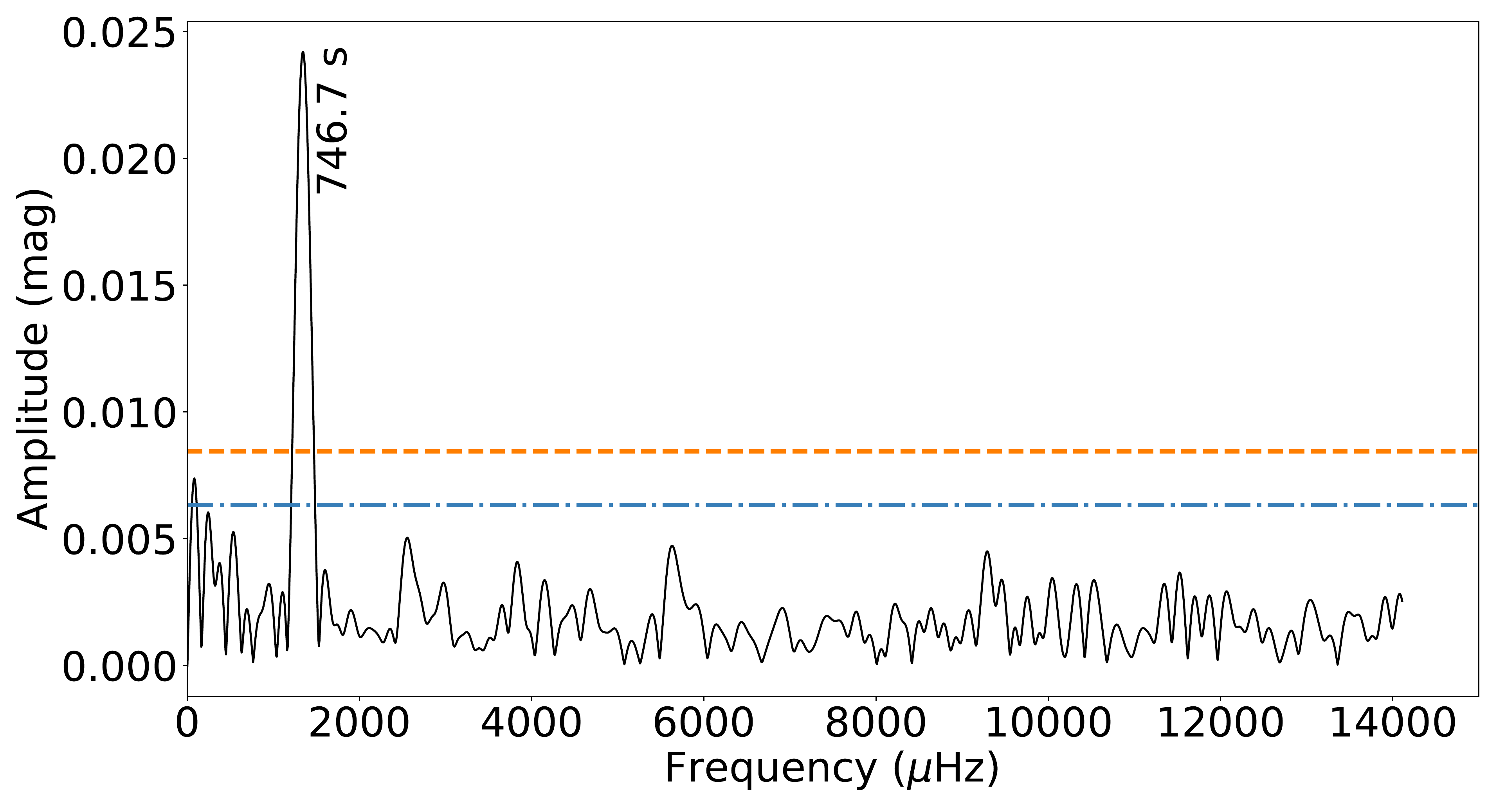}
\caption{Fourier Transform for the three known ZZ Cetis with new detected periods. From top to bottom: BPM 30551, SDSS J092511.63$+$050932.6 and SDSS J215905.53$+$132255.8. Note that the FT shows peaks above the 3$\sigma$ but below the 4$\sigma$ detection limit, adopted in this work. 
\label{FT-known}}
\end{figure}

In this work we performed a follow up of ten known ZZ Ceti stars. For most of them, this is the first time follow--up observations are published since the discovery of their variable nature.
The results from the observations are summarized in Table \ref{known-old}, where we list the frequencies, periods and amplitudes obtained in this work, and the  data reported in previous works (see last column of the table). The FT for the objects for which we found modes with new periods are shown in Figure \ref{FT-known}. In some cases, low amplitude peaks appear in the FT after the pre-whitening process is done. 

\subsection{Possible variables}

\begin{figure}
\includegraphics[width=\columnwidth]{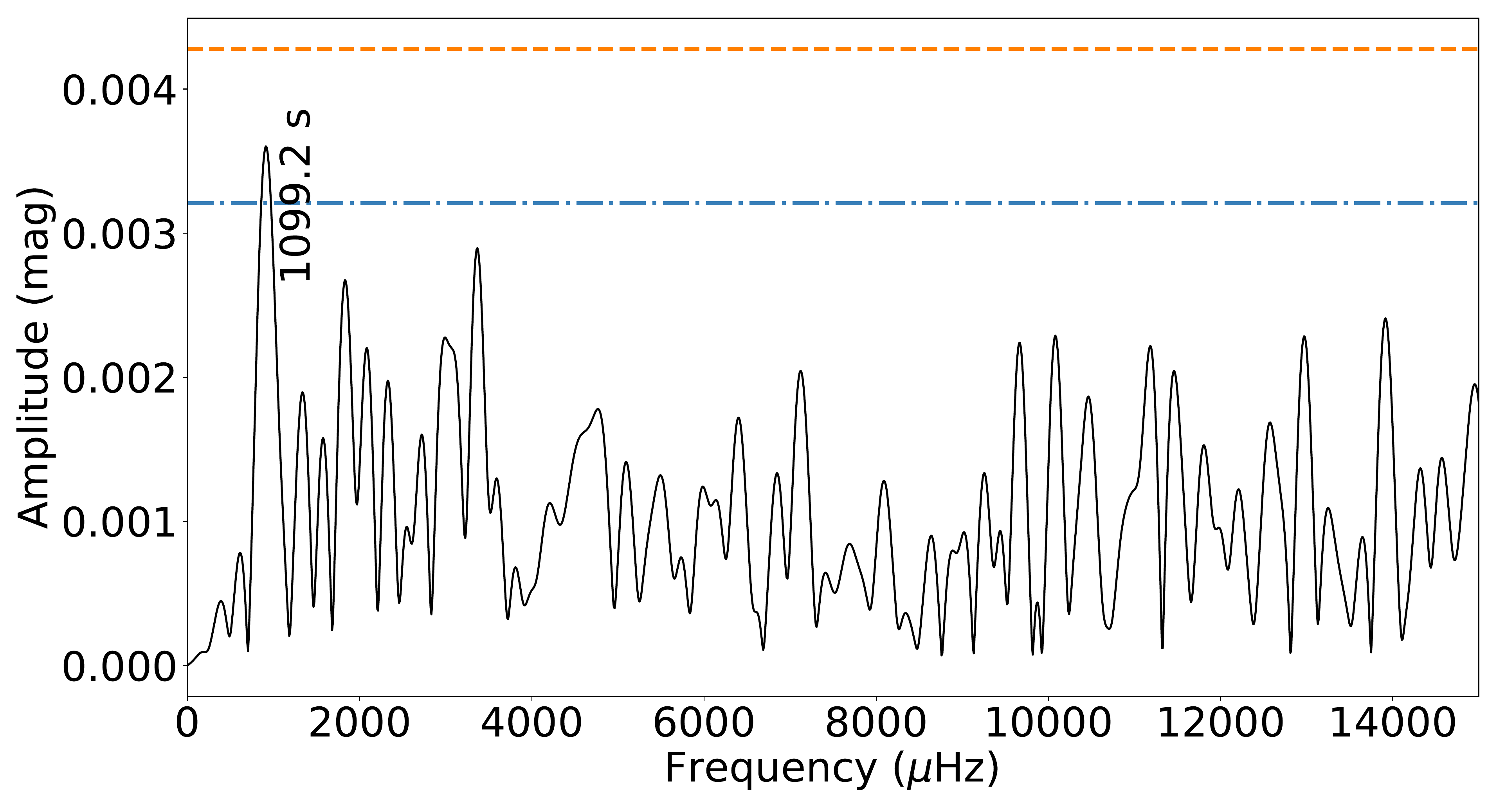}
\includegraphics[width=\columnwidth]{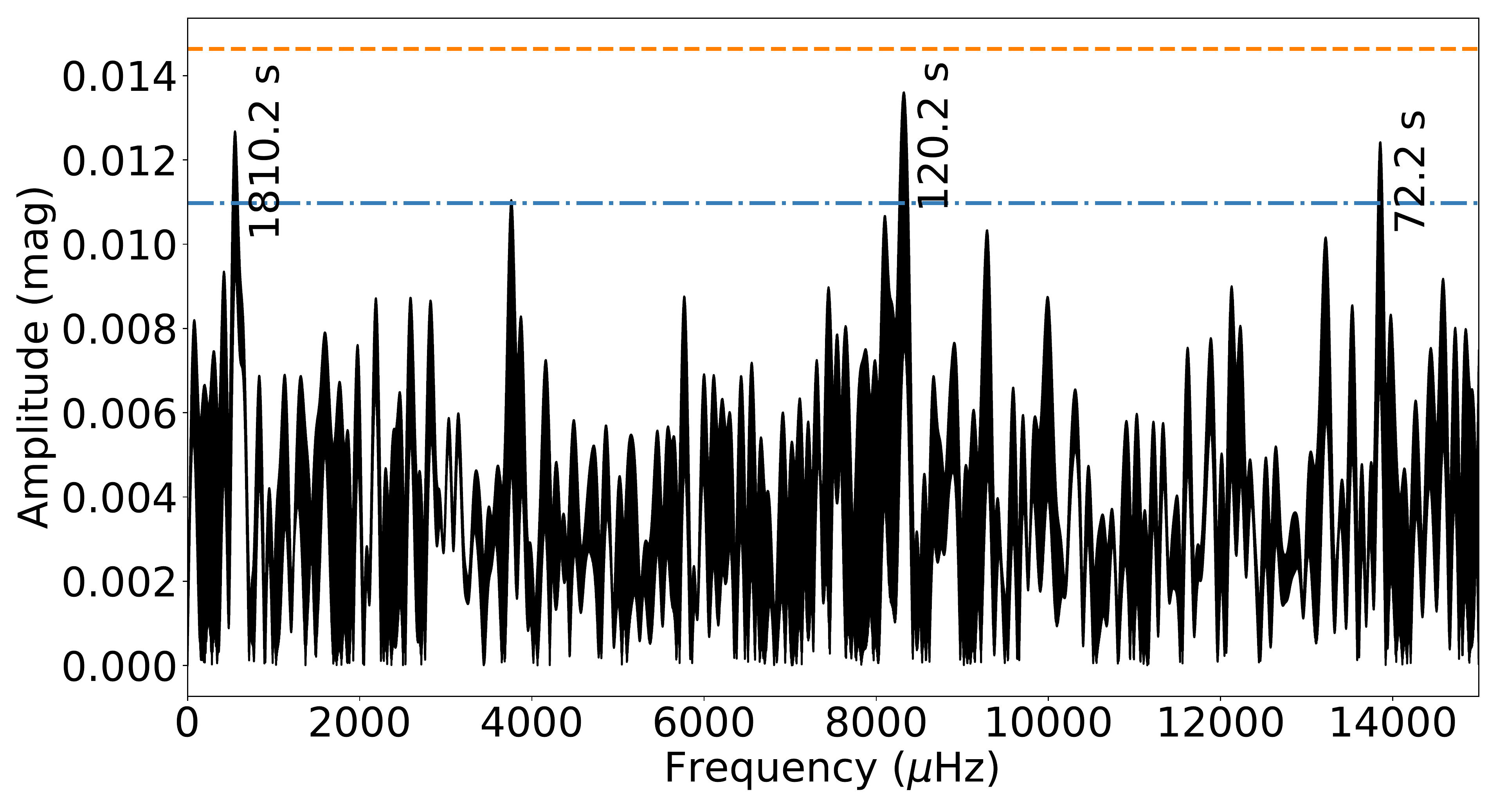}
\includegraphics[width=\columnwidth]{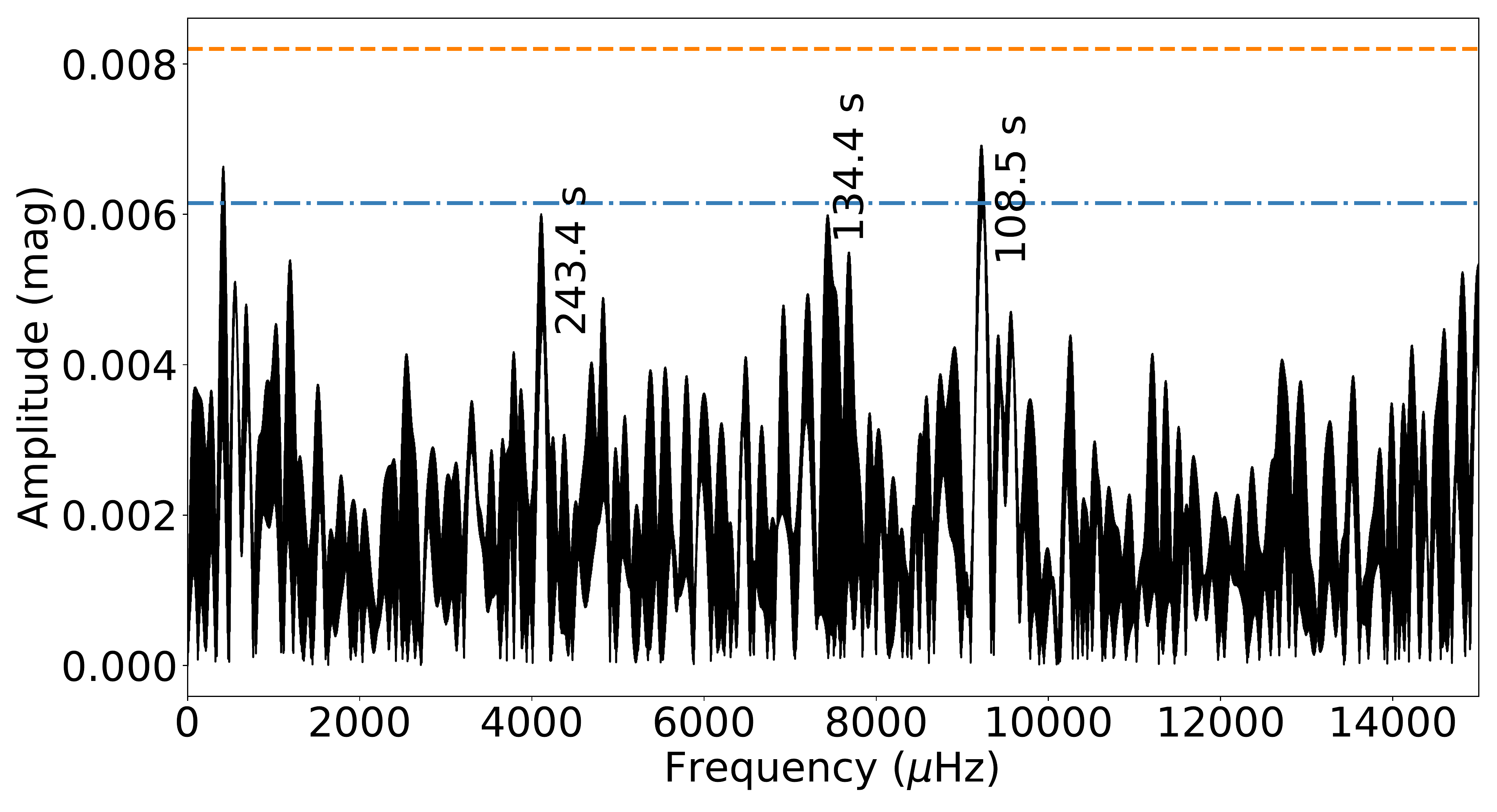}
\includegraphics[width=\columnwidth]{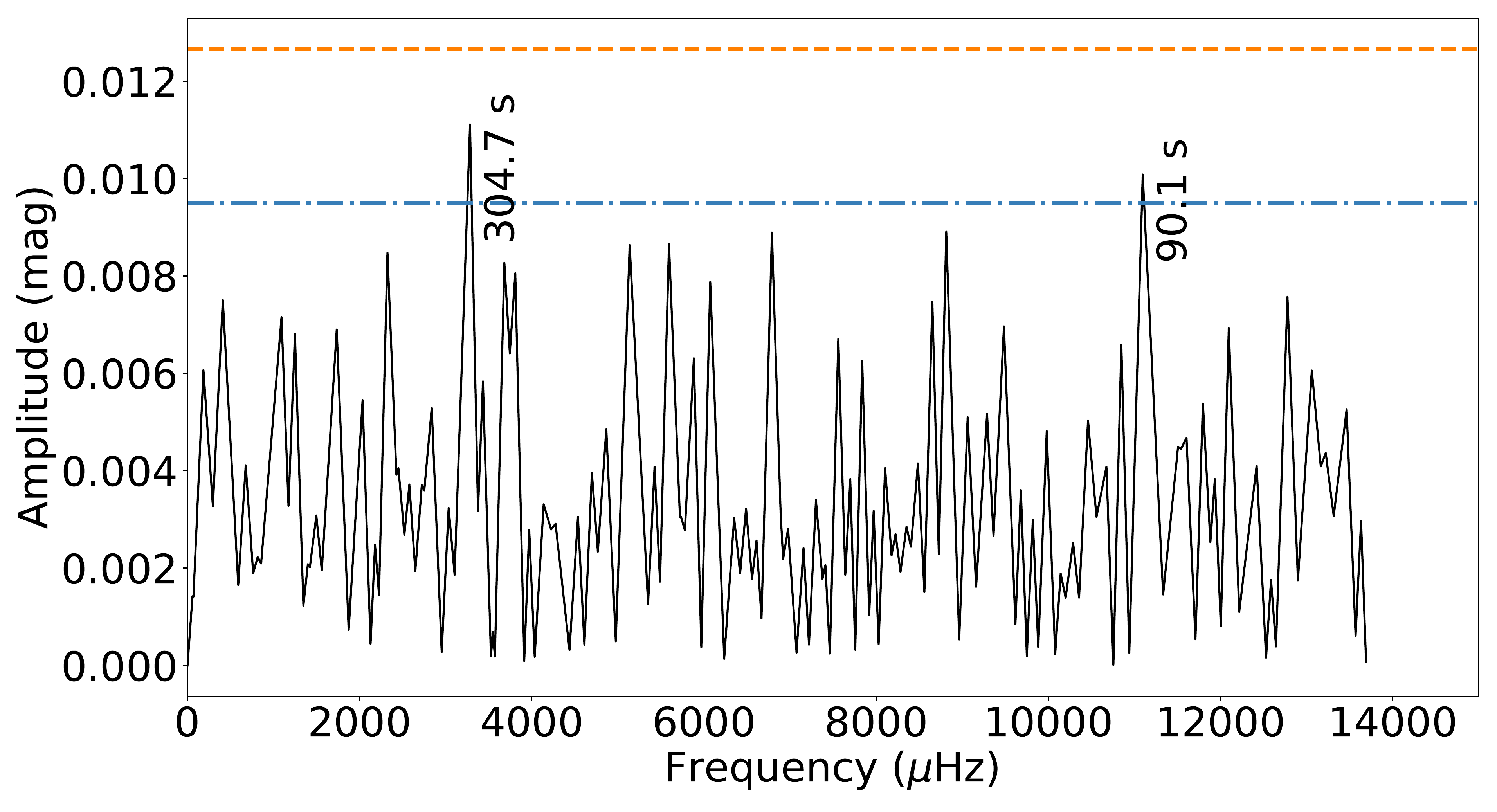}
\caption{Fourier Transform for the four objects classified as possible variables. From top to bottom: LP375$-$51, SDSS J095703.09$+$080504.8, SDSS J212441.27$-$073234.9 and SDSS J213159.88$+$010856.3. Note that the FT shows peaks above the 3$\sigma$ but below the 4$\sigma$ detection limit, adopted in this work. \label{possible}}
\end{figure}

For the candidates LP375$-$51, SDSS J095703.09$+$080504.8, SDSS J212441.27$-$073234.9 and SDSS J213159.88$+$010856.3 (see Table \ref{journal}), we detected variability over the 3$\sigma$ detection limit but below 4$\sigma$ on the FT. In these cases, the S/N was not sufficient to confirm variability and these objects are only classified as candidates. The FT for these objects are shown in Figure \ref{possible}. For LP375$-$51 the FT shows a peak at 1099.2 s. This long period is compatible with the low spectroscopic effective temperature reported for this object. On the other hand, the FT for SDSS J095703.09$+$080504.8 shows two peaks with periods of 120.2 and 72.2 s, compatible with a blue edge pulsator. Similar to SDSS J095703.09$+$080504.8, SDSS J212441.27$-$073234.9 shows a spectroscopic effective temperature characteristic of a blue edge pulsator, and a short period of 108.5 s in the FT. Finally, SDSS J213159.88$+$010856.3 shows one period at 304.7 s, compatible with a warm ZZ Ceti, in agreement with its spectroscopic effective temperature. The second mode with period of 90.1 s is probably instrumental due to the integration time of 30s. We list the periods between 3 and 4$\sigma$ in Table \ref{pi-possible}. Further observations are required to confirm the variable nature of these stars. 

\begin{table}
	\centering
	\caption{Possible variables found in this work, showing peaks between 3 and 4$\sigma$ in the FT. We include the frequency, period and amplitude for each peak, in columns 2, 3 and 4, respectively. In column 5 we list the 4$\sigma$ limit (mma). 
	* Peak with amplitude below $3\sigma$. }
	\label{pi-possible}
	\scalebox{0.95}[0.95]{ \hspace{-7mm}
	\begin{tabular}{lcccc} % four columns, alignment for each
\hline
star & Freq  & Period & Amp  & 4$\sigma$ \\
     & ($\mu$Hz) & (s) & (mma) &\\
\hline
SDSS J095703.09$+$080504.8 & 8321.95 & 120.2 & 12.9 & 14.63\\
             & 552.48  & 1810.02 & 13.5  & \\  
             & 13850.6 & 72.2 & 11.7 & \\
LP 375$-$51 & 909.75 & 1099.2 & 3.6 & 4.28\\      
SDSS J212441.27$-$073234.9 & 9218.58 & 108.5 & 6.9 & 8.20\\
             & 4108.62 & 243.4 & 5.9* & \\
             & 7438.69 & 134.4 & 5.9* & \\  
SDSS J213159.88$+$010856.3 & 3281.46 & 304.7 & 11.11 & 15.89 \\
\hline
	\end{tabular}}
\end{table}

\subsection{NOV}

From the observed sample, we did not detect any variability on the FT for four objects, within the detection limit given by the S/N, and thus they are classified as {\it Not Observed to Vary} (NOV). We list the objects in Table \ref{nov} along with the detection limit from our observations. We recommend a follow up observations given that the detection limit is higher than the typical amplitudes observed in ZZ Cetis, especially near the blue edge of the instability strip \citep[e.g.][]{2013MNRAS.430...50C}.

\begin{table}
	\centering
	\caption{Objects with no detected periodicities. We include the magnitude in the $g$ filter (column 2) and the amplitude of the noise in the FT, as a detection limit (column 3).}
	\label{nov}
	\begin{tabular}{lcc} % four columns, alignment for each
\hline
star & g & 4$\sigma$ (mma)\\
\hline
SDSS J113325.69$+$183934.7 & 17.59 & 8 \\
WD 1454$-$0111 & 17.34 & 10 \\
SDSS J161005.17$+$032356.1 & 18.55 & 7\\
SDSS J235932.80$-$033541.1 & 17.91 & 2 \\
\hline 
	\end{tabular}
\end{table}

\section{Asteroseismological fits}
\label{fits}

In this section we present a detailed asteroseismological analysis of all observed objects, that showed variability. That includes the 10 known ZZ Cetis and the four new variables reported in this work. 
The DA white dwarf models used in this work are the result of full evolutionary computations of the progenitor stars, from the ZAMS, through the hydrogen and helium central burning stages, thermally pulsating and mass loss AGB phase and finally the planetary nebulae domain. They were generated using the {\tt LPCODE} evolutionary code \citep[see][for details]{2010ApJ...717..897A, 2010ApJ...717..183R, 2015MNRAS.450.3708R}. The stellar mass values go from 0.493 $M_{\odot}$ to $1.05 M_{\odot}$, with a hydrogen layer mass in the range of $\sim 4\times 10^{-4} M_*$ to $\sim 10^{-10} M_*$ depending on the stellar mass. 
Non--radial, adiabatic, $g-$mode pulsations were computed using the adiabatic version of the {\tt LP-PUL} pulsation code described in \citet{2006A&A...454..863C}. We employ an extended version of the model grid presented in \citet{2017ApJ...851...60R}, that includes 6 new cooling sequences with stellar masses between 0.5 and 0.7$M_{\odot}$, along with $\sim 8$ hydrogen layer values for each sequence, depending on the mass.

For each object we search for an asteroseismological re\-presentative model, that best matches the observed periods. To this end, we seek for the theoretical model that minimizes the quality function given by \citet{2008MNRAS.385..430C}:

\begin{equation}
S(M_*, M_H, T_{\rm eff}) = \sqrt{\sum_{i=1}^{N}{\rm min}\left[\frac{[\Pi_k^{\rm th}-\Pi_k^{\rm obs}]^2 A_i}{\sum^N_{i=1} A_i}\right]}
\end{equation}  

\noindent where $\Pi^{\rm th}$ is the theoretical period that better fits the observed $\Pi^{\rm obs}$, and the amplitudes $A_i$ are used as weights for each period. In this way, the period fit is more influenced by those modes with large observed amplitudes. 

The results of the asteroseismological fits are presented in Tables \ref{fit1} and \ref{fit2}, corresponding to the new variables and the known variables, respectively. For each object we list the effective temperature, stellar mass and thickness of the hydrogen envelope for the seismological model, in columns 2, 3 and 4 respectively. Column 5 shows the value of the observed period while the theoretical periods are listed in column 6 along with the harmonic degree (col 7) and the radial order (col 8). Finally, the value of the quality function $S$ is listed in column 9. The first model listed is the one we choose to be the best--fit model for that particular object.

In Table \ref{tabla-sismology} we list the structural parameters of the asteroseismological models selected as best--fit models for each star analysed in this paper. The uncertainties for $M_*$, $T_{\rm eff}$ and $\log(L/L_{\odot})$ were computed by using the following expression \citep{1986ApJ...305..740Z, 2008MNRAS.385..430C}:

\begin{equation}
\sigma_i^2 = \frac{d_i^2}{(S-S_0)}
\end{equation}

\noindent where $S_0= S(M_*^0, M_H^0, T_{\rm eff}^0)$ is the minimum of the quality function $S$ reached at $ (M_*^0, M_H^0, T_{\rm eff}^0)$, and $S$ is the value of the quality function when we change the parameter $i$ by an amount $d_i$, keeping fixed the other parameters. The uncertainties in the remaining quantities are derived from the uncertainties in  $M_*$, $T_{\rm eff}$ and $\log(L/L_{\odot})$. These uncertainties represent the internal errors of the fitting procedure.

\begin{table*}
\caption{Structural parameters for the best fit models corresponding to each DAV  star analysed in this paper. The uncertainties are the internal errors of the fitting procedure.}
\centering
\scalebox{0.9}[0.9]{ \hspace{-7mm}
\begin{tabular}{lcccccccc}
\hline
     Star     &   $\log g$       &  $T_{\rm eff}$ [K]  &  $M_*/M_{\odot}$   & $M_{\rm H}/M_*$ &     $M_{\rm He}/M_*$   & $\log(L/L_{\odot})$ & $\log(R/R_{\odot})$ & $X_{\rm O}$\\
\hline
 BPM 30551    & $ 8.08\pm 0.03 $ & $11\,578\pm 65 $ & $0.632\pm 0.014$ & $ 4.65\times 10^{-9}$ & $1.75\times 10^{-2} $ & $-2.632\pm 0.010$ & $-1.921\pm 0.011$ & $0.755$\\
SDSS J092511.63$+$050932.6 & $ 8.13\pm 0.02 $ & $11\,385\pm 54 $ & $0.675\pm 0.016$ & $ 1.34\times 10^{-5}$ & $ 7.65\times 10^{-3}$ & $-2.680\pm 0.018$ & $-1.931\pm 0.008$ & $0.707$\\
HS 1249$+$0426 & $ 8.02\pm 0.04 $ & $11\,564\pm 95 $ & $0.609\pm 0.016$ & $ 1.41\times 10^{-5}$ & $2.45\times 10^{-2}$ & $-2.594\pm 0.014$ & $-1.901\pm 0.014$ & $0.723$\\
WD1345$-$005 & $ 8.14\pm 0.06 $ & $11\,676\pm 196 $ & $0.686\pm 0.011$ & $ 4.40\times 10^{-5}$ & $9.27\times 10^{-3}$ & $-2.646\pm 0.030$ & $-1.935\pm 0.028$ & $0.718$\\
HE 1429-037  & $ 7.91\pm 0.03 $ & $11\,404\pm 44 $ & $0.548\pm 0.005$ & $ 5.33\times 10^{-5}$ & $4.20\times 10^{-2} $ & $-2.545\pm 0.007$ & $-1.868\pm 0.012$ & $0.697$\\
SDSS J161218.08$+$083028.1 & $ 8.46\pm 0.05 $ & $12\,312\pm 366 $ & $0.878\pm 0.041$ & $ 2.85\times 10^{-6}$ & $2.59\times 10^{-3} $ & $-2.758\pm 0.051$ & $-2.037\pm 0.017$ & $0.611$\\
GD 385       & $ 8.33\pm 0.08 $ & $12\,147\pm 196 $ & $0.800\pm 0.037$ & $ 4.05\times 10^{-6}$ & $4.74\times 10^{-3} $ & $-2.700\pm 0.028$ & $-2.000\pm 0.035$ & $0.648$\\
SDSS J215905.53$+$132255.8 & $ 8.52\pm 0.08 $ & $11\,771\pm 169 $ & $0.917\pm 0.040$ & $ 3.89\times 10^{-6}$ & $1.31\times 10^{-3} $ & $-2.879\pm 0.024$ & $-2.058\pm 0.029$ & $0.609$\\
SDSS J221458.37$-$002511.9 & $ 8.46\pm 0.06 $ & $11\,568\pm 124 $ & $0.878\pm 0.041$ & $ 4.12\times 10^{-8}$ & $2.59\times 10^{-3} $ & $-2.874\pm 0.018$ & $-2.042\pm 0.020$ & $0.648$\\
SDSS J235040.72$-$005430.9 & $ 8.17\pm 0.03 $ & $10\,061\pm 85 $ & $0.690\pm 0.015$ & $4.49\times 10^{-8}$ & $7.67\times 10^{-3} $ & $-2.934\pm 0.015$ & $-1.950\pm 0.011$ & $0.684$\\
SDSS J082804.63$+$094956.6 & $ 8.08\pm 0.07 $ & $11\,502\pm 204 $ & $0.646\pm 0.014$ & $ 1.38\times 10^{-5}$ & $1.48\times 10^{-2} $ & $-2.640\pm 0.031$ & $-1.921\pm 0.032$ & $0.742$\\
SDSS J094929.09$+$101918.8 & $ 8.18\pm 0.03 $ & $11\,460\pm 96 $ & $0.705\pm 0.016$ & $ 1.30\times 10^{-5}$ & $7.65\times 10^{-3} $ & $-2.705\pm 0.015$ & $-1.951\pm 0.015$ & $0.661$\\
GD 195       & $ 8.17\pm 0.05 $ & $12\,206\pm 99 $ & $0.705\pm 0.016$ & $ 2.58\times 10^{-5}$ & $7.64\times 10^{-3} $ & $-2.589\pm 0.014$ & $-1.946\pm 0.023$ & $0.661$\\
L495$-$82    & $ 8.00\pm 0.03 $ & $10\,798\pm 60 $ & $0.593\pm 0.016$ & $ 4.58\times 10^{-6}$ & $2.39\times 10^{-2} $ & $-2.705\pm 0.010$ & $-1.899\pm 0.012$ & $0.704$\\
%J2124-0732   & $ 7.84\pm 0.09 $ & $11\,929\pm 452 $ & $0.525\pm 0.014$ & $ 1.88\times 10^{-4}$ & $--- \times 10^{-3} $ & $-2.426\pm 0.067$ & $-1.845\pm 0.049$ & $---$\\

\hline
\label{tabla-sismology}
\end{tabular}}
\end{table*}

%From our sample of DA white dwarf stars, four candidates were found to be variable.  We will present in the following sections a detail analysis of the asteroseimological fit for each object.

\subsection{New ZZ Cetis}

In this section we describe in detail the asteroseismological fits for the four new ZZ Ceti stars discovered in this paper. The results are presented in Table \ref{fit1}.

\begin{table*}
	\centering
	\caption{Best fit model for the four new ZZ Ceti stars. The effective temperature, stellar mass and the mass of the hydrogen envelope are listed in columns 2, 3 and 4, respectively. We list the observed periods used in the asteroseismological fit in column 5. The theoretical periods, harmonic degree and radial order are listed in columns 6, 7 and 8, respectively. The value of the quality function $S$ in seconds is listed in column 9.}
	\label{fit1}
	\begin{tabular}{lcccccccc} 
			\hline
Star & $T_{\rm eff}$ & Mass ($M_{\odot}$) & $\log (M_H/M_{\odot})$ & $\Pi_{obs}$ & $\Pi_{Th}$ & $\ell$ & $k$  &  $S$ (s) \\
\hline
SDSS J082804.63$+$094956.6 & 11502 & 0.646 & -4.86  &  285.79  &  286.55  &  1  &  4  & 0.94  \\
             &       &       &        &  196.31  &  195.04  &  1  &  2  &       \\
             &       &       &        &  255.78  &  256.31  &  1  &  3  &       \\
             & 11620 & 0.686 & -5.52  &  285.79  &  285.37  &  1  &  4  & 0.428 \\
             &       &       &        &  196.31  &  196.38  &  1  &  2  &       \\
             &       &       &        &  255.78  &  256.21  &  2  &  7  &       \\             
 \hline 
SDSS J094929.09$+$101918.8   & 11460 & 0.705 & -4.86  &  199.31  &  199.33  &  1  &  2  & 0.0025 \\
%             & 11529 & 0.560 & -5.36  &  199.31  &  199.31  &  1  &  2  & 0.0002      \\
\hline 
GD 195       & 12206 & 0.705 & -4.59  &  465.16  &  464.92  &  1  &  9  & 0.255 \\  
             &       &       &        &  649.20  &  648.93  &  1  & 14  &       \\
%             & 11121 & 0.745 & -9.24  &  465.16  &  465.09  &  1  &  7  & 0.126 \\
             &       &       &        &  649.20  &  649.37  &  1  & 11  &       \\
\hline               
L495$-$82    & 10798 & 0.593 & -5.34  & 365.16  & 368.87 & 2 & 10 & 2.547 \\ 
             &       &       &        & 581.22  & 578.55 & 1 &  9 & \\
             &       &       &        & 711.94  & 710.27 & 2 & 22 & \\
             &       &       &        & 803.72  & 803.08 & 1 & 14 & \\
             &       &       &        & 902.42  & 905.19 & 1 & 16 & \\
             &       &       &        & 1008.59 & 1005.43 & 1 & 18 & \\
             &       &       &        & 1100.28 & 1100.88 & 1 & 20 & \\
             & 11630 & 0.632 & -7.35  & 365.16 & 365.80 & 2 & 10 & 3.013 \\
             &       &       &        & 581.22 & 575.94 & 1 &  9 &  \\
             &       &       &        & 711.94 & 707.29 & 1 & 12 & \\
             &       &       &        & 803.72 & 807.22 & 1 & 14 & \\
             &       &       &        & 902.42 & 903.66 & 1 & 16 & \\
             &       &       &        & 1008.59 & 1003.99 & 1 & 18 & \\
             &       &       &        & 1100.28 & 1099.80 & 2 & 36 & \\
\hline                   
%J2124$-$0732 & 11929 & 0.525 & -3.73 & 130.20  & 130.157 & 1 & 1 & 0.043\\
%             & 11384 & 0.525 & -3.70 & 130.20  & 130.202 &  1 & 1 & 0.0017 \\
%             & 11924 & 0.493 & -3.50 & 130.20  & 130.206 & 1 & 1 & 0.006 \\
%\hline   
 	\end{tabular}
\end{table*}

\subsubsection{SDSS J082804.63$+$094956.6}

The new ZZ Ceti SDSS J082804$+$094956.6 shows three periods, with the mode at 285.79 s having the largest amplitude. This star shows period pulsations, shorter than 350 s, so we expect it to be close to the blue edge of the instability strip. However, the 3D corrected spectroscopic effective temperature of $11\, 691 \pm 53$ K is closer to the middle of instability strip for the stellar mass of SDSS J082804.63$+$094956.6. 
The results from the seismological fit are listed in Table \ref{fit1}. The first model corresponds to a fit with all modes with $\ell=1$,  while for the second we allowed the mode with 255.78 s to be either $\ell=1$ or 2. Both models are characterized by an effective temperature around $11\, 600$ K, in agreement with the spectroscopic value. The hydrogen envelope is thinner than the canonical value, but is still considered a thick envelope.

\subsubsection{SDSS J094929.09$+$101918.8}

This star shows one period above the 4$\sigma$ detection limit, with a period of 199.31 s, thus we consider this period for our seismological fit. Since we only have one period we need to make some additional restrictions to obtain a theoretical representative model. From spectroscopy, \citet{2019MNRAS.486.2169K}  obtained a 3D corrected effective temperature and surface gravity of $11\, 685 \pm 65$ K and $\log g = 8.073 \pm 0.034$, leading to a stellar mass of $0.664\pm 0.027 M_{\odot}$. %However, using the distance, magnitude and color from {\it Gaia}, we obtain an effective temperature of $\sim 11\, 000$ K and an stellar mass of $\sim 0.5-0.6 M_{\odot}$, where we use the models from Bergeron for the {\tt Gaia} filters. 
The photometric temperature obtained from the SDSS filters \citep[see][for details on the procedure]{2019MNRAS.482..649O} is $11\, 700 \pm 187$ K with $\log g = 8.00 \pm 0.1$ in agreement with the spectroscopy from \citet{2019MNRAS.486.2169K}.

For our seismological fit we consider only the mode with a period of $\sim$ 199 s, which is the only one with an amplitude larger than 4$\sigma$ in the FT. 
The results from our seismological fit are listed in Table \ref{fit1}. The solution is characterized by a stellar mass of  $0.705 M_{\odot}$ and a thick hydrogen envelope. The best fit model for SDSS J094929.09$+$101918.8 has a period of 292.31 s, with $\ell =2$ and $k=9$ that can fit the mode with a period 291.20 s.

\subsubsection{GD 195} From our observations we find two pulsation modes for GD 195, with periods of 465 s and 649 s. For these period values we expect the star to be a warm ZZ Ceti, with effective temperature around $\sim 11\, 500$ K, located in the middle of the instability strip. Since the modes show similar amplitudes in the FT, we consider that both have the same harmonic degree. In this case we expect a degeneracy in the solutions, and we need to use an additional restriction, which in this case can be the spectroscopic temperature and mass. 
The seismic solution compatible with the spectroscopic determinations is characterized by an effective temperature near the blue edge of the instability strip. The solution also shows a thick hydrogen envelope, considering that the stellar mass is 0.705$M_{\odot}$ (see Table \ref{fit1} for details). A second solution, with a lower value of the quality function, is found when we relax the condition on the effective temperature. The stellar mass is somewhat larger but the effective temperature is $\sim 11\, 000$ K, closer to the red edge of the instability strip. Also, the hydrogen envelope mass is the thinnest of our model grid for this stellar mass. A lower effective temperature is compatible with the observed periods, being larger than $\sim 350$ s \citep{2006ApJ...640..956M}. In addition, a low effective temperature is compatible with the colors from {\it Gaia} for this object leading to an effective temperature of $\sim 11\, 000$ K (see Section \ref{gaia}).     

\subsubsection{L495$-$82} L495$-$82 is a rich pulsator with seven detected modes. This is compatible with its low effective temperature of $11\, 029 \pm 160$ K. We consider seven periods in our seismological fit, as shown in Table \ref{fit1}. As an additional restriction, we consider the mode with the largest amplitude, and a period of 902.42 s, to be $\ell=1$. We obtain a best fit model with a stellar mass of 0.593 $M_{\odot}$ and a low effective temperature, compatible with the values from spectroscopic and {\it Gaia} colors (see Section \ref{gaia}). The hydrogen envelope, corresponds to a thick envelope. 
Since the star shows a period of 365.16 s, we consider it to be too short for a pulsator near the red edge of the instability strip, and more characteristic of warm ZZ Ceti, with an effective temperature of $\sim 11\, 500$ K \citep{2006ApJ...640..956M}. With this consideration, we found a second minimum of the quality function characterized by an effective temperature of $\sim 11\, 600$ K. However, the hydrogen envelope is a factor of 100 thinner than the previous model.

\subsection{Known variables}
\label{known}

\begin{table*}
	\centering
	\caption{Best fit model for the known ZZ Cetis, using the list of observed modes (see text for details). The effective temperature, stellar mass and the mass of the hydrogen envelope are listed in columns 2, 3 and 4, respectively. We list the observed periods used in the asteroseismological fit in column 5. The theoretical periods, harmonic degree and radial order are listed in columns 6, 7 and 8, respectively. The value of the quality function $S$ in seconds is listed in column 9.}
	\label{fit2}
	\begin{tabular}{lcccccccc} 
			\hline
Star & $T_{\rm eff}$ & Mass ($M_{\odot}$) & $\log (M_H/M_{\odot})$ & $\Pi_{obs}$ & $\Pi_{Th}$ & $\ell$ & $k$  &  $S$ (s) \\
\hline
BPM 30551 & 11578 & 0.632 & -8.33  & 460.06  &  459.97  &  2  & 13 & 1.86  \\
          &       &       &        & 649.35  &  647.17  &  1  & 10 & \\
          &       &       &        & 775.23  &  772.91  &  1  & 13 & \\
          &       &       &        & 831.03  &  832.45  &  1  & 14 & \\
          &       &       &        & 959.78  &  958.68  &  2  & 30 & \\
          &       &       &        & 986.36  &  987.60  &  2  & 31 & \\       
\hline                      
SDSS J092511.63$+$050932.6 & 11385 & 0.675 & -4.87 & 1127.10 & 1127.01 & 1 & 24 & 1.89 \\
             &       &       &       & 1159.00 & 1163.08 & 1 & 25 & \\
             &       &       &       & 1255.84 & 1255.74 & 1 & 27 & \\
             &       &       &       & 1341.00 & 1339.61 & 1 & 29 & \\
             & 11241 & 0.705 & -7.35 & 1127.10 & 1127.93 & 2 & 38 & 0.64 \\
             &       &       &       & 1159.00 & 1159.12 & 2 & 39 & \\
             &       &       &       & 1255.84 & 1255.09 & 1 & 24 & \\
             &       &       &       & 1341.00 & 1341.49 & 1 & 26 & \\
\hline                       
HS 1249$+$0426 & 11564 & 0.609 & -4.85 & 294.89 & 294.90 & 1 & 4 & 0.001 \\
\hline
WD1345$-$0655 & 11676 & 0.686 & -4.36 & 195.2 & 194.94 & 1 & 2 & 0.22 \\
              &       &       &       & 254.4 & 254.47 & 1 & 3 & \\                        
  \hline            
HE 1429$-$037  & 11404 & 0.548 & -4.27 & 450.10 & 452.39 & 1 & 7 & 1.29 \\
              &       &       &       & 821.74 & 821.03 & 1 & 15 & \\ 
              &       &       &       & 969.00 & 969.70 & 2 & 33 & \\
              &       &       &       & 1084.90 & 1083.92 & 1 & 21 & \\
\hline
SDSS J161218.08$+$083028.1  & 12312 & 0.878 & -5.54 & 115.122 & 115.187 & 1 & 1 & 0.033\\
              & 12619 & 0.686 & -4.36 & 115.122 & 115.123 & 1 & 1 & 0.001 \\
%              & 11906 & 0.917 & -5.77 & 115.122 & 115.118 & 1 & 1 & 0.004\\
\hline                  
GD 385        & 12147 & 0.800 & -5.39 & 127.93 & 127.53 & 1 & 1 & 0.21 \\
              &       &       &       & 256.09 & 256.14 & 1 & 4 & \\
              & 11560 & 0.646 & -6.34 & 127.93 & 127.79 & 2 & 2 & 0.22 \\
              &       &       &       & 256.09 & 256.31 & 1 & 3 & \\  
\hline
SDSS J215905.53$+$132255.8  & 11771 & 0.917 & -5.41 & 683.70 & 684.39 & 1 & 18 & 0.58 \\
              &       &       &       & 746.67 & 746.32 & 2 & 35 & \\
              &       &       &       & 801.00 & 800.97 & 2 & 38 & \\ 
              & 11688 & 0.976 & -6.46 & 683.70 & 684.71 & 1 & 17 & 1.09\\
              &       &       &       & 746.67 & 746.19 & 2 & 33 & \\
              &       &       &       & 801.00 & 801.51 & 2 & 36 & \\     
\hline 
SDSS J221458.37$-$002511.9  & 11568 & 0.878 & -7.38 & 195.08 & 195.61 & 1 & 2 & 0.22 \\
              &       &       &       & 255.20 & 255.10 & 1 & 4 & \\
              & 11605 & 0.686 & -4.36 & 195.08 & 195.50 & 1 & 2 & 0.24 \\
              &       &       &       & 255.20 & 254.87 & 1 & 4 & \\  
\hline
SDSS J235040.72$-$005430.9  & 10061 & 0.690 & -7.35 & 271.87 & 272.93 & 1 & 3 & 0.99 \\
              &       &       &       & 304.74 & 303.67 & 1 & 4 & \\
              &       &       &       & 390.32 & 391.08 & 1 & 5 & \\
              & 10290 & 0.660 & -7.33 & 271.87 & 272.12 & 2 & 6 & 0.31 \\
              &       &       &       & 304.74 & 304.55 & 1 & 4 & \\
              &       &       &       & 390.32 & 389.93 & 1 & 5 & \\
\hline                           
 	\end{tabular}
\end{table*}

We present the asteroseismological fits for the known ZZ Ceti stars that were observed in this work. For the fit we consider all the periods observed for each object, listed in the columns 2 and 4 of Table \ref{known-old}. 
When a detected frequency is close to one previously detected by other authors, we consider the uncertainties in the frequency to determine whether it is a new mode or not. 
The results of our seismological fit for the known ZZ Cetis are listed in Table \ref{fit2}. We present the fitting process for each object below. 

{\bf BPM 30551:} BPM 30551 was observed by \citet{1976ApJ...209..853H}. Several periods were detected between $\sim 300$ and $\sim 2300$ s in the 10 nights. In previous seismological studies, only two periods were used, with 606.8 and 744.7 s \citep{2009MNRAS.396.1709C, 2013ApJ...779...58R,2012MNRAS.420.1462R}. For our seismological fit we consider the six modes detected in this work, with periods between 460 and  986 s. As a result we find a best fit model characterized by a stellar mass of 0.632 $M_{\odot}$ and a thin envelope with $\sim 5\times 10^{-9} M_*$.  

{\bf SDSS J092511.63$+$050932.6:} This star is one of the coolest ZZ Cetis, with a spectroscopic effective temperature less than $\sim 11\, 000$ K. From the FT we detected one period of 1247.5 s. Considering the uncertainty in the frequency for this period, of 198 $\mu$Hz, we consider it to be the same mode as the one with a period of 1264.3 s, detected by \citet{2010MNRAS.405.2561C}, with a difference of $\delta \nu = 10.6 \mu$Hz between both determinations. For our seismological fit, we consider the mean frequency, corresponding to a period of 1255.84 s, along with the other three periods detected in previous works. If we fix the harmonic degree to be $\ell=1$ for all modes we obtain a representative model with $0.675 M_{\odot}$ and a thick hydrogen envelope. By relaxing this condition, two periods are fitted with  quadrupole ($\ell =2$) mode. The solution has a larger mass and a thinner hydrogen envelope, possibly due to the core--envelope symmetry \citep{2003MNRAS.344..657M}.

{\bf HS 1249$+$0426:} This object shows only one peak in the FT at 294.91 s. \citet{2006AA...450.1061V} also detected one period of 288.9 s. Given the uncertainty in the frequency, we concluded that they are the same period, and use the the value obtained in this work for the seismological fit. We consider only $\ell =1$ modes in our fit. The solution is similar to that found by \citet{2012MNRAS.420.1462R}, characterized by a canonical stellar mass and an effective temperature of $\sim 11\, 500$ K. Finally, the hydrogen envelope is a factor of three thinner than the previous fit, but still considered a thick envelope.

{\bf WD1345$-$0055:} \citet{2004ApJ...607..982M} reported the detection of two short periods for WD1345$-$0055. From our observations we recover the one period of 195.2 s. For our seismological fit, we consider the two modes. As a result we found a representative model with a stellar mass of $0.686 M_{\odot}$ and a canonical envelope, that predicted by single stellar evolution for this stellar mass. Both modes are fitted with theoretical dipole ($\ell=1$) mode.

{\bf HE 1429$-$037:} For this object, \citet{2005AA...443..195S} reported the detection of four periods between 450 and 1084 s. From our observations we found one mode with a period of 821.74 s. Considering the uncertainties we conclude that this period corresponds to the period of 829.3 s, detected by \citet{2005AA...443..195S}. For the seismological fit we consider the mean frequency, corresponding to a periods of 825.505 s. The seismic solution has a low stellar mass of $0.548 M_{\odot}$ and a thick hydrogen envelope of $ 1.9 \times 10^{-5} M_*$.

{\bf SDSS J161218.08$+$083028.1}: \citet{2013MNRAS.430...50C} reported the detection of two short periods, part of a triplet with a central component with a period of $\sim 115$ s. We recover these periods from our observations, with the additional possible detection of a period of 112.09 s, which is part of the triplet. We consider the spectroscopic determinations of mass and $T_{\rm eff}$ as additional restrictions in our seismological fit, since we only have one observed mode. The seismic solution has a high stellar mass and effective temperature, as expected from a short period pulsator, with a hydrogen envelope of $2.85\times 10^{-6} M_*$. If we relax the restriction in stellar mass, we found a second solution with a $0.686 M_{\odot}$ and a thicker envelope.  

{\bf GD 385}: GD 385 is a hot ZZ Ceti showing two modes. We recover both modes from our observations and did not detected new periodicities. For our seismological fit, first we fixed the harmonic degree to $\ell=1$ for both modes and obtained a hot solution with a stellar mass of $0.8M_{\odot}$, somewhat larger than the spectroscopic mass (see Table \ref{Tlogg}). The second solution presented in Table \ref{fit2}, was obtained by fixing the mode  with the largest amplitude to be a dipole mode and letting the harmonic degree for the second mode free. The solution shows a stellar mass compatible with the spectroscopy but the effective temperature is low, as compared to other pulsators that show a period $\sim 195$ s. 
%This solution is similar to that found by \citet{2012MNRAS.420.1462R} using a similar grid.

{\bf SDSS J215905.53$+$132255.8:} This object is the most massive ZZ Ceti analysed in this work. Two pulsation modes were reported by \citet{2005ApJ...625..966M}, with periods of 683.7 and 801.0 s. In this work we find a period of 746.67 s with a large amplitude, and a second period with 678.8 s after subtracting the main peak from the FT. The second period has a frequency that is $\delta \nu = 10\mu$Hz from the frequency corresponding to the mode with 683.7 s previously reported. Thus we consider that they are the same mode and use three periods in our seismological fit. 
The model that minimized the quality function is characterized by a stellar mass of 0.917 $M_{\odot}$ as it is shown in Table \ref{fit2}. We also consider a second solution, closer to the one obtained by \citet{2013ApJ...779...58R} using two periods. In this case, the stellar mass is 0.976$M_\odot$ and the hydrogen envelope is $\sim 10$ times thinner than the first solution, which is related to the core-envelope symmetry  \citep{2003MNRAS.344..657M}. In this case the core should be 7\% crystallized.

%Also, the period 746.67 s is close to a linear combination $(f_1 + f_2) / 2$ of the periods presented in \citet{2005ApJ...625..966M}, with a difference of $\delta \nu = 16.25 \mu$Hz, which is also within the uncertainties. The results from the fit do not change if only two periods are considered in our analysis. 

{\bf SDSS J221458.37$-$002511.9:} For this object, we recover one of the two periods presented by \citet{2005ApJ...625..966M}, with a period of $\sim 255$ s. For our seismological fit we use the two known periods. We find two representative theoretical models with similar quality functions, listed in Table \ref{fit2}. The first model has a stellar mass of $0.878M_{\odot}$ and a thin hydrogen envelope, while the second solution is characterized by a stellar mass of $0.686 M_{\odot}$ and a thick envelope.  Both models fit the observed modes with $ \ell=1$ modes and show effective temperatures of $\sim 11\, 600$ K, in agreement with the spectroscopy.

{\bf SDSS J235040.72$-$005430.9:} This ZZ Ceti is an ultra--cool ZZ Ceti, with an spectroscopic effective temperature of $\sim 10\, 600$ K. From our observations we recover three modes, presented in \citet{2004ApJ...607..982M}. We carried two seismological fits, one fixing the harmonic degree to be $\ell=1$ for all modes, and a second by considering that the mode with the highest amplitude is a $\ell=1$ mode while leaving the harmonic degree free for the remaining two modes. Both fitting procedures lead to a cool solution with a thin hydrogen envelope $\log (M_H/M_*) \sim -7.3$. 

This object is very odd in the sense that the effective temperature is very low as compared with the bulk of ZZ Ceti stars. \citet{2013ApJ...779...58R} considered that this object, and other ultra--cool ZZ Cetis, could be low mass white dwarfs, with stellar masses below $0.3 M_{\odot}$, which is in line with the mass obtained from parallax (see  Section \ref{gaia}). Other explanation include the possibility of a binary companion, in which case, the determination of the spectroscopic mass being affected by the presence of the companion \citep{fuchs2018}. This hypothesis will be explored in an future paper.

\begin{figure}
\includegraphics[width=0.6\textwidth]{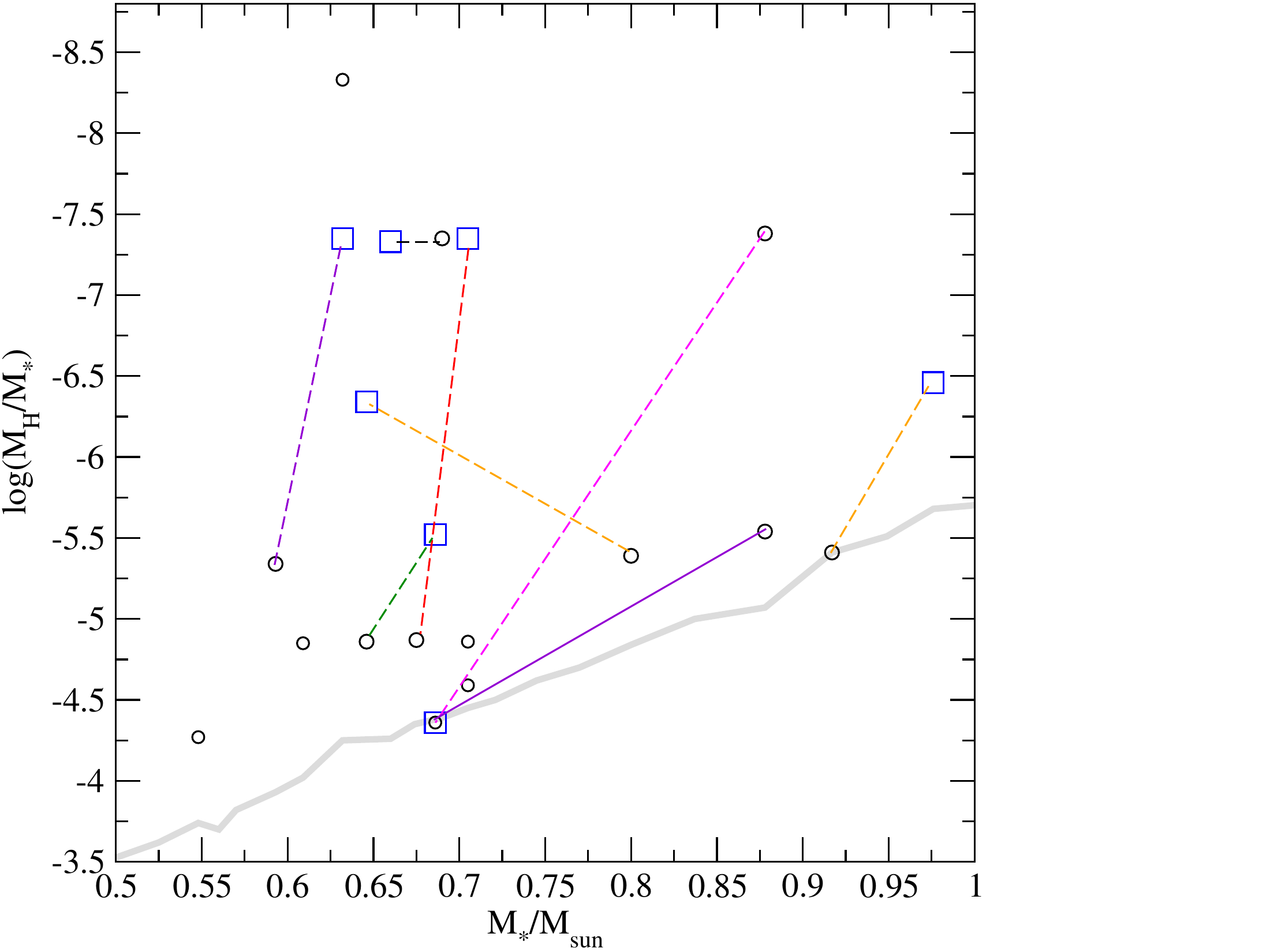}
\caption{Values of hydrogen envelope mass in terms of the stellar mass, corresponding to all the asteroseismological models of the 14 objects analized in this work. Black circles and blue squares correspond to the first and second  solution, respectively (see Tables \ref{fit1} and \ref{fit2}). Solutions correspondign to the same object are conected with a line. The thick gray line depicts the canonical values of the hydrogen envelope thickness \citep{2017ApJ...851...60R}\label{extra}}
\end{figure}

%----------------------------------------------------------------------------
To summarize, in Figure \ref{extra} we plot all the seismological solutions listed in Tables \ref{fit1} and \ref{fit2}, in the stellar mass -- thickness of the hydrogen envelope plane. With black circles, we plot
the best-fit models for each star, whereas blue squares represent the second solutions, when present. Solutions corresponding to the same object
are joint together with a line. The thick, gray line indicates
the high limit of the hydrogen mass, as predicted by stellar
evolution. Note that for several objects, we obtain two possible
seismological solutions, even after additional restrictions are considered. Usually one is characterized by a higher stellar
mass and a thin hydrogen envelope and other characterized
by a lower mass and a thicker hydrogen layer. For example, 
for L495$-$82 we obtained a best-fit model characterized
by $M_∗ = 0.593 M_{\odot}$ and $\log(M_H /M _∗) = -5.34$ and a second
solution with $M_∗ = 0.632 M_{\odot}$ and $\log(M_H /M _∗) = -7.35$. This degeneracy in solutions is related to the so called ``core--envelope symmetry'' discussed in \citet{2003MNRAS.344..657M}, where a
sharp feature in the Brunt--V\"ais\"al\"a frequency in the envelope can produce the same period changes as a bump placed in the core.

 %-------------------------------------------------------------------
 
\section{Using {\it Gaia} data}
%\section{Thing with {\tt Gaia}}
\label{gaia}

Using the data from the {\it Gaia} mission, we have additional information on the ZZ Cetis. From the distance and magnitudes we can estimate the stellar mass and effective temperature, independently from the spectroscopy. 
Using hydrogen rich atmosphere models for {\it Gaia} magnitudes \citep[see][for details]{2019MNRAS.486.2169K} combined with mass--radius relation from \citet{2019MNRAS.484.2711R}, we transform absolute magnitude $M_G$ and color $G_{bp} - G_{rp}$ into stellar mass and effective temperature. The absolute magnitude is computed from the apparent magnitude and the distance. For stellar masses lower than $0.5 M_{\odot}$ we use the atmosphere models from the Montreal Group (P. Bergeron, private communication, \citep[see also][]{2011ApJ...737...28B}.  Note that the uncertainties in the effective temperature are underestimated since the magnitude filter from the {\it Gaia} satellite are quite broad, in the case of white dwarf stars.

The results are summarized in Table \ref{gaia-data}. We list parallax, distance, $G$ apparent magnitude and color  $G_{bp} - G_{rp}$ in columns 2, 3, 4 and 5, respectively. The distance was taken from \citet{2018AJ....156...58B}, except for the objects marked with an asterisk, for which we compute the distance from the inverse of the parallax. Since for all objects the uncertainties in the parallax is less than 5\%, we do not expect large deviations \citep{2018AJ....156...58B}. 
Also listed are the absolute magnitude $M_G$ (col 6) and the stellar mass (col 7) and effective temperature (col 8) computed in this work. In the last column, we specify the status of the star, as known variable, new variable, possible variable and NOV.

\begin{table*}
	\centering
	\caption{{\tt Gaia} data for all observed targets. We list the parallax (col 2), distance in pc (col 3), apparent $G$ magnitude (col 4) and colour (col 5), along with the absolute magnitude $M_G$ (col 6) and the stellar mass (col 7) and effective temperature (col 8) computed in this work (see text for details). The last column indicates the status of the object from this work.
	 * Distances computed by taking the inverse of the parallax angle.}
	\label{gaia-data}
	\scalebox{0.92}[0.92]{ \hspace{-7mm}
	\begin{tabular}{lcccccccc}
	\hline
star & parallax (mas) & distance (pc) & $G$ &  $G_{bp} - G_{rp}$ & $M_G$ & Mass ($M_{\odot}$) & $T_{\rm eff}$ & Class\\ 
\hline
BMP 30551   & $20.027 \pm 0.014$ & $49.860 \pm 0.080$ & 15.477 & 0.027 & 11.985 & $0.6372\pm 0.0057$ & $11\,106\pm 70$ & known\\
SDSS J092511.63$+$050932.6  & $24.663 \pm 0.061$ & $40.499 \pm 0.100$ & 15.271 & 0.054 & 12.231 & $0.7117\pm 0.0061$ & $10\,831\pm 58$ & known\\
%PG1149+058  & $24.787 \pm 0.065$ & $40.296 \pm 0.105$ & 14.980 & 0.041 & 11.951 & 0.608$ & known\\
HS 1249$+$0426 & $14.648 \pm 0.068$ & $68.139 \pm 0.316$ & 16.045 & 0.007 & 11.874 & $0.6184\pm 0.0072$ & $11\,409\pm 72$ & known\\
WD1345$-$0055 & $9.820 \pm 0.105$ & $101.552 \pm 1.092$ & 16.789 & -0.005 & 11.750 & $0.5881\pm 0.0117$ & $11\,533\pm 125$ & known\\
HE 1429$-$037  & $14.341 \pm 0.105$ & $69.598 \pm 0.514$ & 16.033 & 0.040 & 11.816 & $0.5602\pm 0.0134$ & $10\,889\pm 162$ & known\\
SDSS J161218.08$+$083028.1  & $7.662 \pm 0.183$ & $130.512 \pm 3.118*$ & 17.831 & -0.025 & 12.253 & $0.8173\pm0.0280$ & $12\,062\pm 248$ & known\\
GD 385      & $21.115 \pm 0.037$ & $47.295 \pm 0.082$ & 15.149 & 0.014 & 11.772 & $0.5751\pm 0.0008$ & $11\,247\pm 60$ & known\\
SDSS J215905.53$+$132255.8  & $5.150 \pm 0.308$ & $194.254 \pm 11.981$ & 18.999 & 0.027 & 12.558 & $0.8074\pm 0.1032$ & $10\,831\pm 970$ & known\\
SDSS J221458.37$-$002511.9  & $7.011 \pm 0.211$ & $142.270 \pm 4.309$ & 17.923 & 0.025 & 12.516 & $0.7177\pm 0.0398$ & $11\,227\pm 364$ & known\\
SDSS J235040.72$-$005430.9  & $4.665 \pm 0.265$ & $214.023 \pm 12.373$ & 18.121 & 0.170 & 11.465 & $0.2998\pm 0.0283$ & $9\,370\pm 180$  & known\\
SDSS J082804.63$+$094956.6  & $6.460 \pm 0.181$ & $154.307 \pm 4.355$ & 17.710 & 0.037 & 11.761 & $0.5310\pm 0.0431$ & $11\,027\pm 283$ & new \\
SDSS J094929.09$+$101918.8  & $6.959 \pm 0.168$ & $143.235 \pm 3.485$ & 17.580 & 0.031 & 11.793 & $0.5686\pm 0.0431$ & $11\,067\pm 288$ & new\\
GD 195        & $9.441 \pm 0.192$ & $105.672 \pm 2.168$ & 16.632 & 0.038 & 11.508 & $0.4459\pm 0.0164$ & $10\,700\pm 100$ & new\\ 
L495$-$82     & $42.779 \pm 0.043$ & $23.375 \pm 0.024*$ & 13.764 & 0.027 & 11.920 & $0.6158\pm 0.0023$ & $11\,106\pm 30$ & new\\
SDSS J095703.09$+$080504.8  & $8.767 \pm 0.166$ & $113.749 \pm 2.172$ & 17.704 & 0.035 & 12.418 & $0.7943\pm 0.0343$ & $11\,067\pm 308$ & possible\\
LP 375$-$51   & $19.758 \pm 0.054$ & $50.537 \pm 0.139$ & 15.701 & - & 12.180 & - & - &  possible\\
SDSS J212441.27$-$073234.9  & $4.165\pm 0.324$  & $240.069\pm 20.221*$ & 18.575 & 0.070 & 11.673 & $0.4710\pm 0.0629$ & $10\,350\pm 100$ & possible\\ 
SDSS J213159.88$+$010856.3  & $5.034 \pm 0.199$ & $198.627 \pm 7.882*$ & 18.403 & 0.064 & 11.912 & $0.5687\pm 0.0323$ & $10\,520\pm 140$ & possible\\
SDSS J113325.09$+$183934.7  & $10.079 \pm 0.278$ & $99.056 \pm 2.757$ & 17.595 & 0.099 & 12.612 & $0.7958\pm 0.0448$ & $10\,338\pm 320$ & NOV\\
WD1454$-$0111 & $9.536 \pm 0.159$ & $104.870 \pm 1.749*$ & 17.343 & -0.049 & 12.239 & $0.8337\pm 0.0325$ & $12\,413\pm 335$ & NOV\\ 
SDSS J161005.7$+$030256.1  & $4.437 \pm 0.239$ & $225.093 \pm 12.452$ & 18.550 & 0.018 & 11.785 & $0.5850\pm 0.0512$ & $11\,307\pm 345$ & NOV\\
SDSS J235932.80$-$033541.1  & $3.254 \pm 0.245$ & $306.690 \pm 23.747$ & 17.910 & -0.035 & 10.472 & $0.2470\pm 0.0247$ & $11\,340\pm 240$ & NOV\\
\hline            
	\end{tabular}}
\end{table*}

\begin{figure}
\includegraphics[width=\columnwidth]{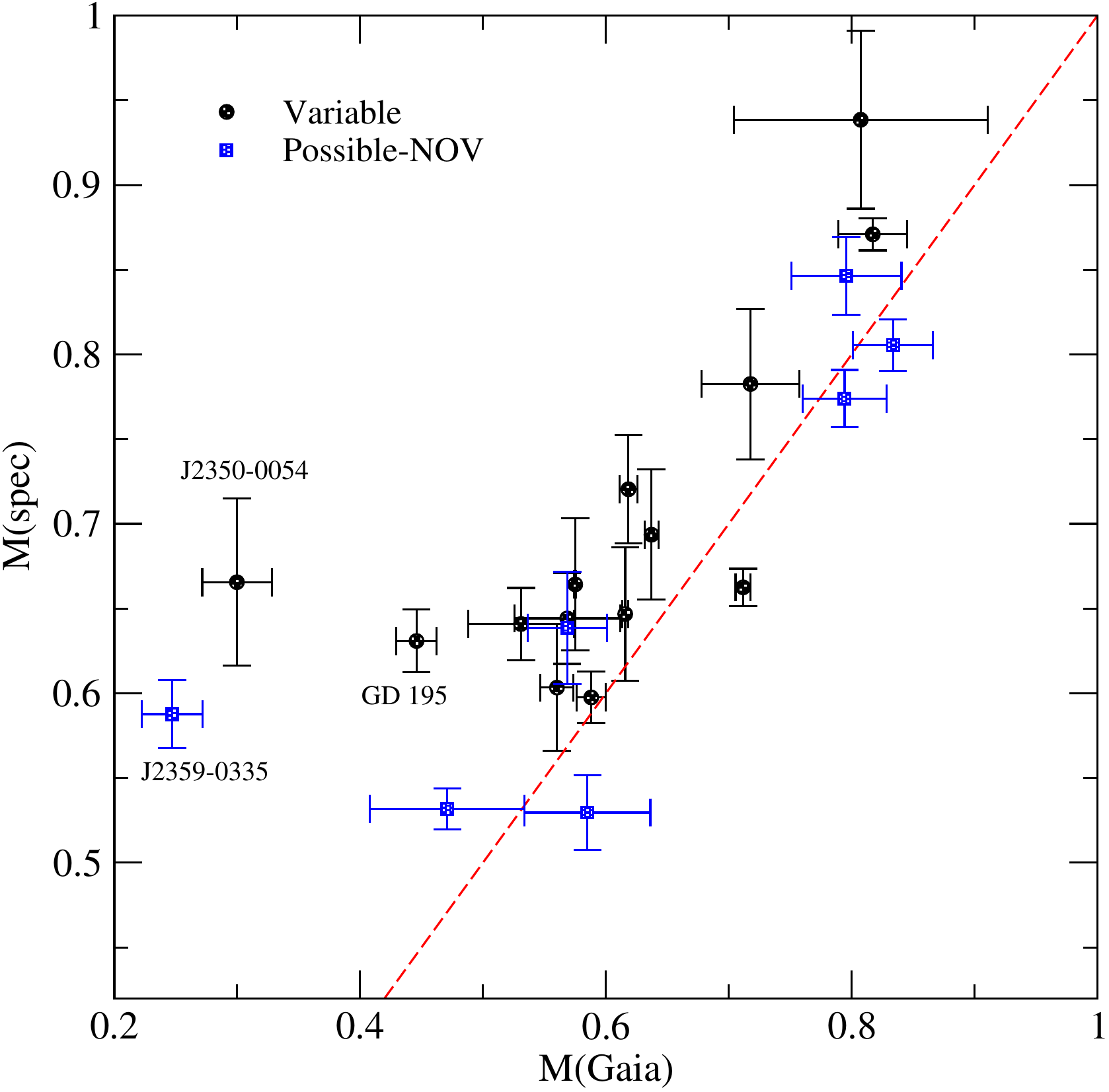}
\caption{Comparison between the values of the stellar mass according to {\it Gaia} data and spectroscopy for all observed targets. Variable white dwarf stars, new and known variables, are depicted with black circles, while the objects not classified as variables, possible and NOV, are depicted with blue squares. The uncertainties are the internal uncertainties of the fitting procedure. The red line represents the 1:1 correspondence. }
\label{masses-gaia-spec}
\end{figure}

\begin{figure}
\includegraphics[width=\columnwidth]{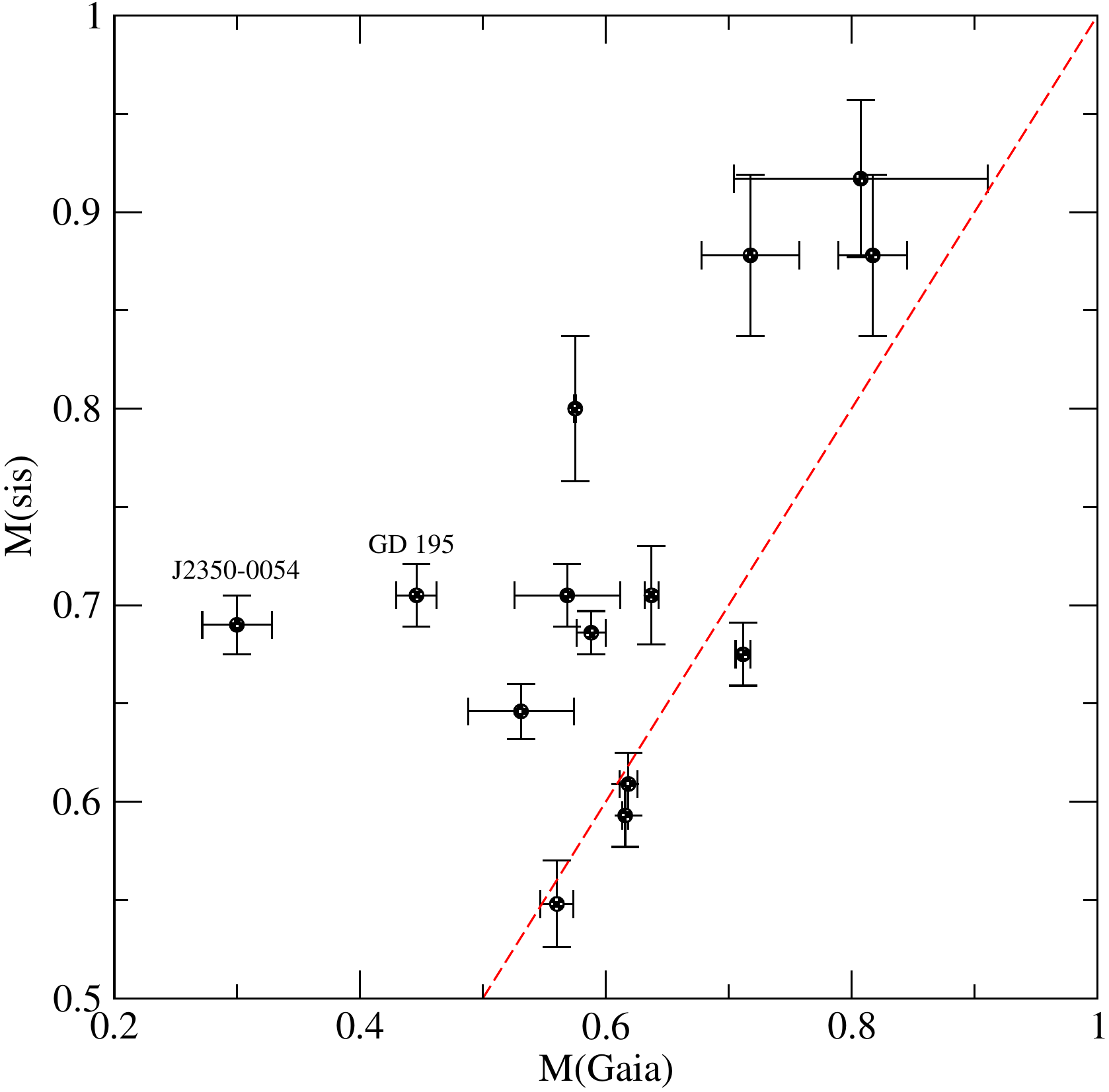}
\caption{Comparison between the values of the stellar mass according to {\it Gaia} data and asteroseismology for all observed targets showing confirmed photometric variability. The uncertainties are the internal uncertainties of the fitting procedure. The red line represents the 1:1 correspondence. }
\label{masses-gaia-sis}
\end{figure}

We compare the stellar mass obtained from distance and {\it Gaia} magnitudes (Table \ref{gaia-data}) with the determinations obtained from spectroscopic values of $\log g$ and effective temperature (Table \ref{atmosphere}) and the seismological mass (Table \ref{tabla-sismology}). Since the evolutionary models used to obtain a seismological representative model for each object are the same that we used to derive the spectroscopic mass from the observed spectroscopic parameters and to determine the mass--radius relation for the atmosphere models, this comparison is worth doing. The results are depicted in Figures \ref{masses-gaia-spec} and \ref{masses-gaia-sis}. 

The comparison between the stellar mass based on {\it Gaia} data and spectroscopy is presented in Figure \ref{masses-gaia-spec}. The variable DA white dwarfs are depicted with black circles, while the objects with no confirmed variability are depicted with blue squares. The uncertainties are the internal uncertainties of the fitting procedure. For most objects, the correspondence between both determinations is not in good agreement, specially for three objects: SDSS J235040.72$-$005430.9, GD 195 and SDSS J235932.80$-$035541.1. In these cases, the stellar mass based on {\it Gaia} data is that of a low mass white dwarf, with stellar masses of 0.2998, 0.4459 and 0.2470 $M_{\odot}$ respectively (see Table \ref{gaia-data} for details). In particular, SDSS J235040.72$-$005430.9 has been a mystery since the discovery of its variability by \citet{2004ApJ...607..982M}, showing an spectroscopic temperature characteristic of the red edge and short pulsation periods, characteristic of the blue edge of the instability strip.  As was mentioned in section \ref{known}, this object can indeed be part of a WD$+$WD binary system, where the flux is dominated by the less massive, brighter component \citep{fuchs2018}. Given this evidence, it is possible that GD 195, and specially SDSS J235932.80$-$035541.1 are also part of an unresolved double degenerate binary system. 

A similar trend is found when we compare the stellar mass based in {\it Gaia} data and the seismological mass obtained from our fits. Since SDSS J235932.80$-$035541.1 is classified as NOV it is not depicted in this figure. As expected, SDSS J235932.80$-$035541.1 lays above the 1:1 correspondence line, with a seismological mass of $0.69 M_{\odot}$. The same happens for GD 195, with a seismological mass of $0.705 M_{\odot}$.

\section{Analysis of the sample}
\label{astero-sample}

In this section we analyse the main results of a sample of $\sim 91$ ZZ Cetis with asteroseismological fits. We include the results from previous asteroseismological fits that used the same grid of models, to be consistent with the results obtained in this work.  From the works of  \citet{2012MNRAS.420.1462R, 2013ApJ...779...58R} and \citet{2017ApJ...851...60R} we selected 77 objects. Finally, we include the 14 ZZ Cetis analysed in this work, with 10 previously known variables and the 4 new ZZ Cetis.
In case one object was analysed more than once, we choose the asteroseismological solution from the most recent asteroseismological fit.

\begin{figure}
\includegraphics[width=\columnwidth]{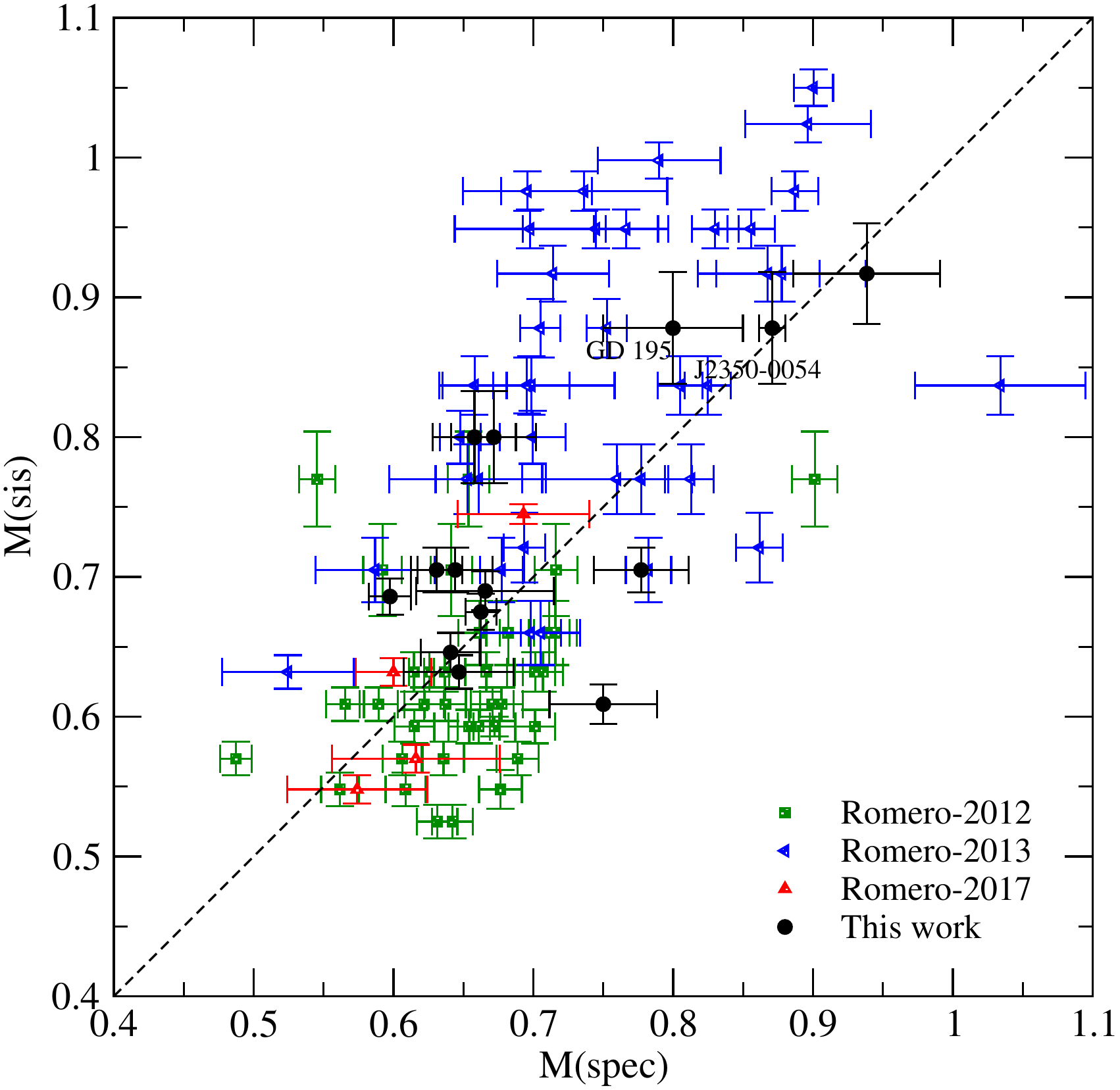}
\caption{Comparison between the values of the stellar mass obtained from spectroscopy with 3D convection correction and asteroseismology for a sample of 91 ZZ Ceti stars.  The sample is taken from \citet{2012MNRAS.420.1462R} (green squares), \citet{2013ApJ...779...58R} (blue triangle-right), \citet{2017ApJ...851...60R} (red triangle-up) and this work (black circle). The uncertainties are the internal uncertainties of the fitting procedure. The dashed line represents the 1:1 correspondence.}
\label{masses}
\end{figure}

In Figure \ref{masses} we compare the stellar mass obtained from spectroscopy and seismology for the sample of 91 ZZ Cetis. The spectroscopic mass is taken from Table \ref{atmosphere}, with 3D convection correction.
The general agreement between both sets of estimates is not quite good, the largest discrepancy being for stellar masses above $\sim 0.75 M_{\odot}$. Note that 3D convection correction in $\log g$ is not completely efficient in the high mass regime \citep{2019MNRAS.482.5222T}, and thus could be the reason for the deviation seen in that mass range. 
However, the bulk of point in Figure \ref{masses} accumulate around the 1:1 correspondence line, demonstrating that no appreciable offset exist between the spectroscopic and asteroseismic estimations of the stellar mass. 
The mean spectroscopic mass for the sample of 91 ZZ Cetis is $\left< M_{\rm spec}\right> = 0.692 M_{\odot}$, $\sim 5\%$ lower than the mean seismological mass for the same sample $\left< M_{\rm sis}\right> = 0.727 M_{\odot}$. Note that these values are largely affected by the 36 massive ZZ Cetis analysed by \citet{2013ApJ...779...58R}, with stellar masses larger than $0.72 M_{\odot}$, affected by the possible shortcoming in the 3D convection correction. Thus, these sample should not be compared with other samples with an homogeneous distribution in stellar mass. If we do not consider the sample from \citet{2013ApJ...779...58R}, we obtain an average spectroscopic mass of $\left< M_{\rm spec}\right> = 0.657 M_{\odot}$, which is only 1.7\% higher than the corresponding mean seismological mass of $\left< M_{\rm sis}\right> = 0.646 M_{\odot}$. 

One of the parameters that can be estimated almost exclusively by asteroseismology is the hydrogen mass left in the envelope of a DA white dwarf star. The value of the hydrogen envelope mass for the objects analysed in this work is listed in column 5 of Table \ref{tabla-sismology}. Note that, depending on the stellar mass, the canonical hydrogen envelopes can have masses ranging from $\sim10^{-3}M_*$, for 0.49$M_{\odot}$ to $10^{-6}M_*$ for $\sim 1M_{\odot}$ \citep{2012MNRAS.420.1462R,2019MNRAS.484.2711R}. 

\begin{figure}
\includegraphics[width=\columnwidth]{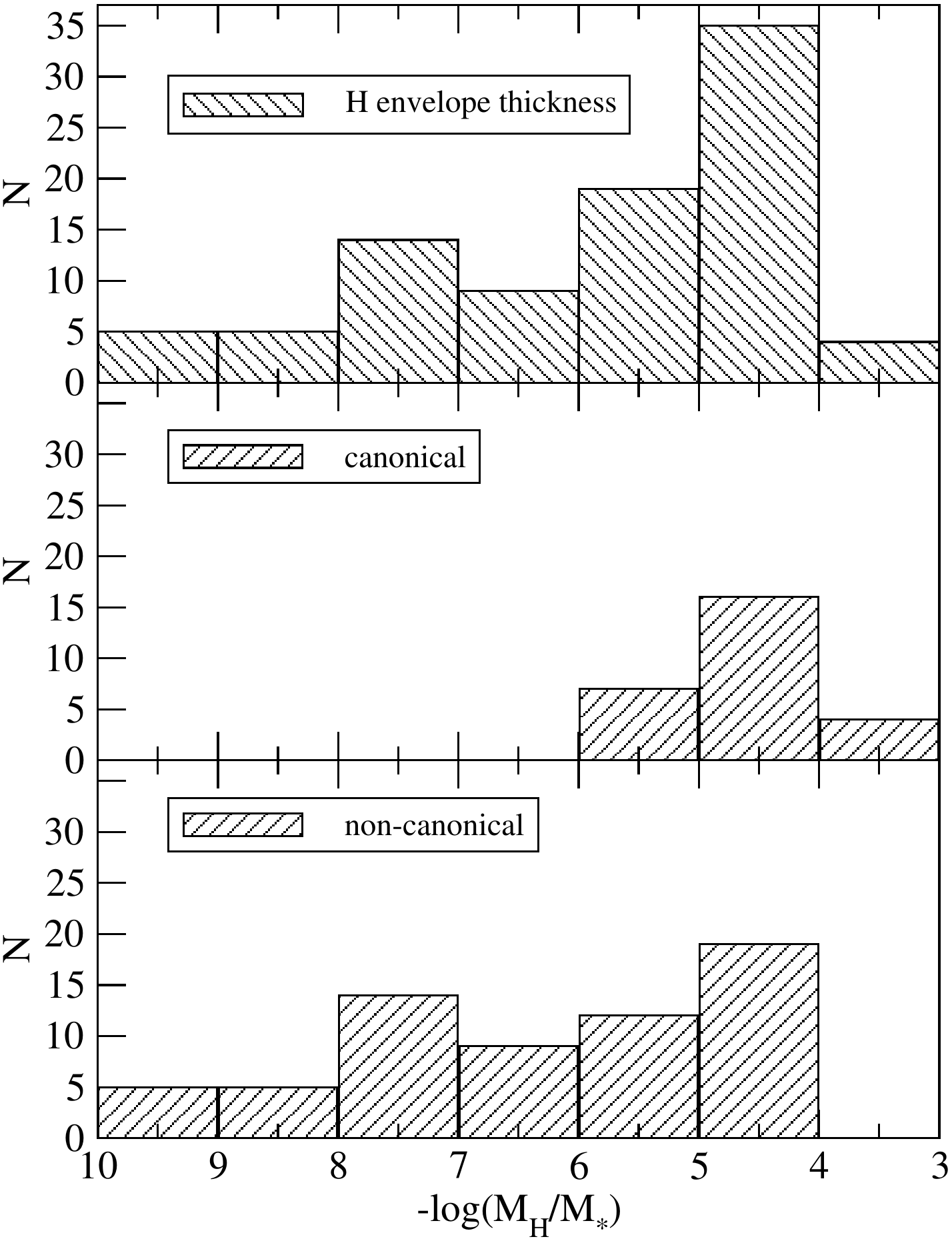}
\caption{ Upper panel: histogram showing the hydrogen envelope thickness distribution for the sample of 77 ZZ Cetis stars. Middle panel: histogram for models with canonical hydrogen envelope thickness, as predicted by canonical evolutionary computations according to the value of the stellar mass. Lower panel: histogram for models with non-canonical envelope thickness. \label{histo}}
\end{figure}

Figure \ref{histo} shows the distribution of the hydrogen envelope thickness for a sample of 91 ZZ Ceti stars (upper panel), taken from \citet{2012MNRAS.420.1462R, 2013ApJ...779...58R, 2017ApJ...851...60R} and this work. The middle and bottom panels show the distribution for the canonical envelopes, those with the thickest envelope allowed by single stellar evolution, and the thin envelopes, respectively. 

From the distribution of hydrogen envelope mass, we note a pronounced maximum of the distribution for $\log(M_H / M_*)$ in the range $-$5 to $-4$, with contributions from both thin envelopes, for the low mass models, and canonical envelopes for masses above $\sim 0.60 M_{\odot}$. A second peak for $\log(M_H/M_*)$ between $-$7 and $-$8 is also present in the distribution, with contributions mainly from the high mass ZZ Cetis \citep{2013ApJ...779...58R}. 
From our sample of 91 ZZ Cetis, we found that 35\% of the best fit models have canonical envelopes, those with the thickest envelope as predicted by single stellar evolution. However, as much as 75\% show hydrogen envelopes thicker than $10^{-6}M_*$ and only 13\% shows very thin hydrogen envelopes with masses below $10^{-8}M_*$. This result is in agreement with the results presented by \citet{2017ASPC..509..255C} from a sample of 16 hot ZZ ceti stars. They found that the best matching models, taken from the model grid presented in \citet{2012MNRAS.420.1462R}, have hydrogen layer masses values at or near the canonically thick limit calculated from nuclear burning, which is consistent with our results.  

The mean value of the hydrogen layer mass is $\left< M_{\rm H}/M_*\right> = 2.3\times 10^{-6}$. This value is $\sim 5$ times larger than that obtained by \citet{2009MNRAS.396.1709C}, with a sample covering a broad range in stellar mass, and using a different model grid. In spite of this difference, both studies conclude that the possible values for the hydrogen envelope span over a large range ($10^{-4} - 10^{-10} M_*$), with a fraction of DA white dwarf stars formed with a hydrogen envelope much thinner than that predicted by single stellar evolution computations. An excellent example of a DA white dwarf with a measured thin hydrogen envelope is 40 Eridani B. \citet{2019MNRAS.484.2711R} obtained a hydrogen mass of $M_{\rm H} = 2.6\times 10^{-8} M_{\odot}$ by comparing the theoretical mass--radius relations for different hydrogen envelope masses with the dynamical stellar mass from  \citet{2017AJ....154..200M} and the radius obtained from photometry and distance \citep{2017ApJ...848...16B}.  

Another evidence of the existence of DA white dwarf with thin hydrogen envelopes was presented by \citet{2019MNRAS.482..649O}, who studied the spectral evolution of white dwarf stars, using a sample of $\sim 13\, 000$ DA and $\sim 3\,000$ non$-$DA white dwarf stars with both spectroscopic data from SDSS DR12 catalogue and the {\it Gaia} DR2 survey.  The authors found that the ratio of non--DA to DA white dwarfs is $\sim 0.075$ for effective temperatures above $22\, 000$ K and increases by a factor of five for effective temperatures cooler than $15\, 000$ K. The most likely explanation for the spectral evolution is the convective mixing of a thin hydrogen envelope into the underlying helium layer of $14\pm 3 \%$ of DA white dwarf stars.

\section{Conclusions}\label{conclusion}

In this work we present the results from ground based observations applied to the search of variable DA white dwarf stars. We report the discovery of four new variables: SDSS J082804.63$+$094956.6, SDSS J094929.09$+$101918.8, GD 195 and L495$-$82. In addition we re-observed 10 known ZZ Cetis to look for new periodicities and to study the stability of the pulsation periods. From the sample of 12 candidates, four objects are classified as possible variables, with peaks in the FT with amplitudes above 3$\sigma$ but below 4$\sigma$, the latter being the detection limit adopted in this work. Our main results are listed below.

The candidates were selected from the SDSS white dwarf catalogue \citep[e.g.][]{2019MNRAS.486.2169K} complemented by the list of DA white dwarfs presented in \citet{2017ApJ...848...11B}. Using the sample of known ZZ Cetis, we selected those objects with spectroscopic atmospheric parameters within the empirical instability strip. 
Since we found new variables among our candidates we believe that this selection method is adequate. Currently, we have $\sim 570$ candidates from the SDSS white dwarf catalogue \citep{2019MNRAS.486.2169K} within the instability strip that have not been studied for variability.

By comparing stellar mass determinations from spectroscopy and seismology with that obtained using {\it Gaia} data, we found three outliers. The stellar mass of SDSS J235040.72$-$005430.9, SDSS J235932.8$-$033541.1 and GD 195 determined using photometry and parallax from {\it Gaia} is 0.299, 0.247 and 0.4459 $M_{\odot}$, respectively, below the stellar mass obtained with spectroscopy, and incompatible with single stellar evolution. Since the lowest mass considered in our model grid is $0.493 M_{\odot}$, the seismological mass, for SDSS J235040.72$-$005430.9 and GD 195 is also higher than the determination obtained with {\it Gaia}. In particular, there is evidence that SDSS J235040.72$-$005430.9 could be an unresolved WD$+$WD system \citep{fuchs2018}, with the flux dominated by the less massive, brighter companion. Thus, within this hypothesis, it is possible that GD 195 and specially SDSS J235932.8$-$033541.1 are also an unresolved double degenerate system. 

Finally, we analyse the properties of a sample of 91 ZZ Ceti stars, that were subject of an asteroseismological study. The distribution of hydrogen envelope mass spans the range $-\log (M_{\rm H}/M_*) = 4-10$, with a pronounced maximum for $\log (M_{\rm H}/M_*)$ between $-4$ and $-5$, in agreement with the results obtained by \citet{2017ASPC..509..255C} based solely on observational data. The mean value for our sample is $\left< M_{\rm H}/M_*\right> = 2.3\times 10^{-6}$. Note that 91 objects correspond to $\sim 36 \%$ of all the ZZ Cetis known to date.

\section*{Acknowledgements}

%We thank the anonymous referee for her/his valuable comments and suggestions. 
ADR, LAA, GO, GRL and SOK acknowledges financial support from CNPq and PRONEX-FAPERGS/CNPq (Brazil). TK acknowledges financial support from Capes (Brazil). DS is supported by the Gemini Observatory, which is operated by the Association of Universities for Research in Astronomy, Inc., on behalf of the international Gemini partnership of Argentina, Brazil, Canada, Chile, the Republic of Korea, and the United States of America. IP  acknowledges funding by the Deutsche Forschungsgemeinschaft under grant GE2506/12-1. 

Based on observations obtained at Observat\'orio do Pico dos Dias/LNA, at the Southern Astrophysical Research (SOAR) telescope, which is a joint project of the Minist\'erio da Ci\^encia, Tecnologia, Inova\c{c}\~ao e Comunica\c{c}\~es (MCTIC) do Brasil, the US National Optical Astronomy Observatory (NOAO), the University of North Carolina at Chapel Hill (UNC), and Michigan State University (MSU) and processed using the Gemini IRAF package, which is operated by the Association of Universities for Research in Astronomy, Inc., under a cooperative agreement with the NSF on behalf of the Gemini partnership: the National Science Foundation (United States), the National Research Council (Canada), CONICYT (Chile), Ministerio de Ciencia, Tecnolog\'ia e Innovaci\'on Productiva (Argentina) and Minist\'erio da Ci\^encia, Tecnologia, Inova\c{c}\~ao e Comunica\c{c}\~es (Brasil). 
This work has made use of data from the European Space Agency (ESA) mission {\it Gaia} (\url{https://www.cosmos.esa.int/gaia}), processed by the Gaia Data Processing and Analysis Consortium (DPAC, \url{https://www.cosmos.esa.int/web/gaia/dpac/consortium}). Funding for the DPAC has been provided by national institutions, in particular the institutions participating in the Gaia Multilateral Agreement. This research  has also made use of the TOPCAT (\url{http://www.starlink.ac.uk/topcat/}) software \citep{2005ASPC..347...29T}, The Montreal White Dwarf Database \citep[][\url{http://www.montrealwhitedwarfdatabase.org/home.html}]{2017ASPC..509....3D}, and NASA Astrophysics Data System.
We thank P. Bergeron for sharing the atmosphere models for DA white dwarf for {\it Gaia} filters. 
The observational data is available upon request to {\url{alejandra.romero@ufrgs.br}.

%the programs SO2016B-016, SO2017A-012 and SO2017B-012 (PI: Alejandra Romero)
%OPD2016A-018, OPD2016B-018, OPD2017A-004, OPD2018A-010 and OPD2018B-007 (PI: Alejandra Romero).

\section*{Appendix}

\begin{table}
\caption{Uncertainties in the measured frequencies and their amplitudes.}
\scalebox{0.8}[0.8]{ \hspace{-7mm}
\begin{tabular}{lccccc}
\hline  
Known    & Freq & $\sigma_{\rm freq}$ & Amp & $\sigma_{\rm amp}$ & Period \\ 
            &($\mu$Hz)  & ($\mu$Hz) & (mma) &  (mma) &(s) \\
            \hline
BPM 30551           & 12033.238 & 0.02    & 11.2  & 0.7  & 831.031   \\
                    & 1289.930  & 5       & 10.2  & 0.7  & 775.235   \\
                    & 1041.895  & 0.02    & 7.7   & 0.6  & 959.789  \\
                    & 2173.602  & 0.02    & 5.7   & 0.5  & 460.065   \\
                    & 1013.831  & 0.08    & 5.5   & 1    & 986.357  \\
                    & 1540.005  & 0.03    & 5.5   & 0.5  & 649.348  \\ 
SDSS J092511.63$+$050932.6 &  801.628  & 13      & 8     & 1    & 1247.5    \\ 
HS 1249$+$0426       & 3391.095  & 9       & 14.5  & 3    & 294.91    \\ 
WD1345$-$0055       & 5121.860  & 84      & 8.9   & 3    & 195.24    \\ 
HE 1429$-$037        & 1216.918  & 35      & 56.93 & 15   & 821.74    \\ 
GD 385              & 3904.829  & 4       & 9.4   & 0.5  & 256.09    \\
                    & 7816.344  & 35      & 3.5   & 0.8  & 127.9     \\ 
SDSS J215905.53$+$132255.8 & 1339.282  & 5       & 24.2  & 1    & 746.7     \\
                    & 1473.210  & 47      & 8     & 2    & 678.8     \\ 
SDSS J221458.37$-$002511.9 & 3920.273  & 4       & 16    & 2    & 255.1     \\ 
SDSS J235040.72$-$005430.9 & 3281.518  & 7       & 18.29 & 2    & 304.7     \\
                    & 2562.004  & 12      & 10.17 & 2    & 390.3     \\
                    & 3678.258  & 57      & 8.2   & 3    & 271.9     \\ 
\hline
New                 &           &          &             &           \\ 
\hline
SDSS J08204.638+094956.6 & 3499.073 & 0.4      & 14.2 & 0.8  & 285.79    \\
                    & 5093.984 & 0.4      & 10.9 & 0.8  & 196.31    \\
                    & 3909.61  & 0.8      & 5.7  & 0.6  & 255.78    \\
SDSS J094929.09+101918.8 & 5017.309 & 0.08     & 3.3  & 0.4  & 199.31    \\
                    & 3434.066 & 0.02     & 1.9  & 0.5  & 291.2     \\
                    & 8403.361 & 0.01     & 2    & 0.4  & 119       \\ 
GD 195               & 2149.798 & 0.2      & 8.7  & 0.1  & 465.16    \\
                    & 1540.375 & 0.2      & 6.2  & 0.1  & 649.2     \\
L495$-$82             & 1108.125 & 13       & 10.48 & 3   & 902.425   \\
                    & 908.856  & 21       & 6.72  & 3   & 1100.283  \\
                    & 1244.21  & 30       & 5.74  & 3   & 803.722   \\
                    & 1404.597 & 10       & 3.92  & 4   & 711.947   \\
                    & 2746.011 & 21       & 3.59  & 4   & 364.164   \\
                    & 991.48   & 51       & 3.74  & 2   & 1008.59   \\
                    & 1720.509 & 38       & 2.62  & 1   & 581.223   \\
                    & 2396.375 & 14       & 2.001 & 0.3 & 417.296   \\
                    & 4578.305 & 17       & 1.41  & 0.3 & 218.421   \\
                    & 2031.562 & 181      & 1.39  & 3   & 492.23    \\
                    & 3018.182 & 121      & 1.21  & 9   & 331.32    \\
                    & 2541.883 & 295      & 1.19  & 5   & 393.409  \\
\hline
Candidate           &                      &              &           \\ 
\hline

SDSS J095703.09$+$080504.8 & 8321.928  & 0.4     & 12.9  & 3    & 120.2     \\
                    & 13850.564 & 2       & 11.7  & 3    & 72.2      \\ 
LP 375$-$51         & 909.750   & 303     & 3.6   & 2    & 1099.2    \\ 
SDSS J212441.27$-$073234.9 & 9218.58   & 0.6     & 6.9   & 2    & 108.5     \\
                    & 4108.62   & 0.7     & 5.9   & 1    & 243.4     \\
                    & 7438.69   & 15      & 5.9   & 2    & 134.4     \\   
SDSS J213159.88$+$010856.3 & 3281.458  & 25      & 11.11 & 4    & 304.7     \\
                    & 11095.163 & 22      & 10.5  & 3    & 90.1     \\
\hline                    
\end{tabular}}
\end{table}

%\begin{figure}
%\includegraphics[width=0.6\textwidth]{online-plot.eps}
%\caption{Values of hydrogen envelope mass in terms of the stellar mass, corresponding to all the asteroseismological models of the 14 objects analized in this work. Black circles and blue squares correspond to the first and second  solution, respectively (see Tables \ref{fit1} and \ref{fit2}). Solutions correspondign to the same object are conected with a line. The thick gray line depicts the canonical values of the hydrogen envelope thickness \citep{2017ApJ...851...60R}\label{extra}}
%\end{figure}

\label{lastpage}

\begin{thebibliography}{}
\bibitem[Althaus et al. (2010)]{2010ApJ...717..897A} Althaus L.~G., C{\'o}rsico A.~H., Bischoff-Kim A., Romero A.~D., Renedo I., Garc{\'\i}a-Berro E., Miller Bertolami M.~M., 2010, ApJ, 717, 897
\bibitem[Baran(2013)]{2013AcA....63..203B} Baran, A.~S.\ 2013, \actaa, 63, 203
\bibitem[B{\'e}dard et al.(2017)]{2017ApJ...848...11B} B{\'e}dard, A., Bergeron, P., \& Fontaine, G.\ 2017, \apj, 848, 11 
\bibitem[Bailer-Jones et al.(2018)]{2018AJ....156...58B} Bailer-Jones C.~A.~L., Rybizki J., Fouesneau M., Mantelet G., Andrae R., 2018, AJ, 156, 58
\bibitem[Bell et al.(2019)]{2019RNAAS...3f..81B} Bell, K.~J., Kosakowski, A., Kilic, M., et al.\ 2019, Research Notes of the American Astronomical Society, 3, 81
\bibitem[Bell et al.(2017)]{2017ASPC..509..303B} Bell, K.~J., Hermes, J.~J., Montgomery, M.~H., et al.\ 2017, 20th European White Dwarf Workshop, 303
\bibitem[Bell, et al.(2015)]{2015ApJ...809...14B} Bell, K.~J., Hermes, J.~J., Bischoff-Kim, A., et al.\ 2015, \apj, 809, 14
\bibitem[Bergeron et al.(2011)]{2011ApJ...737...28B} Bergeron, P., Wesemael, F., Dufour, P., et al.\ 2011, \apj, 737, 28
\bibitem[Bognar \& Sodor(2016)]{2016IBVS.6184....1B} Bognar, Z., \& Sodor, A.\ 2016, Information Bulletin on Variable Stars, 6184, 1
\bibitem[Bond et al.(2017)]{2017ApJ...848...16B} Bond, H.~E., Bergeron, P., \& B{\'e}dard, A.\ 2017, \apj, 848, 16
\bibitem[Brickhill (1991)]{1991MNRAS.251..673B} Brickhill A.~J., 1991, MNRAS, 251, 673
\bibitem[Castanheira et al. (2013)]{2013MNRAS.430...50C} Castanheira B.~G., Kepler S.~O., Kleinman S.~J., Nitta A., Fraga L., 2013, MNRAS, 430, 50
\bibitem[Castanheira et al. (2010)]{2010MNRAS.405.2561C} Castanheira B.~G., Kepler S.~O., Kleinman S.~J., Nitta A., Fraga L., 2010, MNRAS, 405, 2561
\bibitem[Castanheira \& Kepler (2009)]{2009MNRAS.396.1709C} Castanheira B.~G., Kepler S.~O., 2009, MNRAS, 396, 1709
\bibitem[Castanheira \& Kepler (2008)]{2008MNRAS.385..430C} Castanheira B.~G., Kepler S.~O., 2008, MNRAS, 385, 430
\bibitem[Castanheira, et al. (2007)]{2007A&A...462..989C} Castanheira B.~G., et al., 2007, A\&A, 462, 989
\bibitem[Castanheira, et al. (2006)]{2006A&A...450..227C} Castanheira B.~G., et al., 2006, A\&A, 450, 227
\bibitem[Clemens et al.(2017)]{2017ASPC..509..255C} Clemens, J.~C., O'Brien, P.~C., Dunlap, B.~H., et al.\ 2017, 20th European White Dwarf Workshop, 255
\bibitem[Clemens(1993)]{1993BaltA...2..407C} Clemens, J.~C.\ 1993, Baltic Astronomy, 2, 407
\bibitem[C{\'o}rsico et al.(2019)]{2019arXiv190700115C} C{\'o}rsico, A.~H., Althaus, L.~G., Miller Bertolami, M.~M., et al.\ 2019, arXiv e-prints, arXiv:1907.00115
\bibitem[C{\'o}rsico \& Althaus (2006)]{2006A&A...454..863C} C{\'o}rsico A.~H., Althaus L.~G., 2006, A\&A, 454, 863
\bibitem[Dolez \& Vauclair (1981)]{1981A&A...102..375D} Dolez N., Vauclair G., 1981, A\&A, 102, 375
\bibitem[Dufour et al.(2017)]{2017ASPC..509....3D} Dufour, P., Blouin, S., Coutu, S., et al.\ 2017, 20th European White Dwarf Workshop, 509, 3 

\bibitem[Fontaine \& Brassard (2008)]{2008PASP..120.1043F} Fontaine G., Brassard P., 2008, PASP, 120, 1043
\bibitem[Fuchs (2018)]{fuchs2018} Josh Fuchs, 2018, Presentation in the 21st European White Dwarf Workshop, Austin, Texas
\bibitem[Gaia Collaboration et al.(2016)]{2016A&A...595A...1G} Gaia Collaboration, Prusti, T., de Bruijne, J.~H.~J., et al.\ 2016, \aap, 595, A1
\bibitem[Gianninas, Bergeron \& Fontaine (2006)]{2006AJ....132..831G} Gianninas A., Bergeron P., Fontaine G., 2006, AJ, 132, 831
\bibitem[Gianninas, Bergeron \& Ruiz (2011)]{2011ApJ...743..138G} Gianninas A., Bergeron P., Ruiz M.~T., 2011, ApJ, 743, 138
\bibitem[Gilliland et al.(2010)]{2010ApJ...713L.160G} Gilliland, R.~L., Jenkins, J.~M., Borucki, W.~J., et al.\ 2010, \apjl, 713, L160
\bibitem[Goldreich \& Wu (1999)]{1999ApJ...511..904G} Goldreich P., Wu Y., 1999, ApJ, 511, 904
%\bibitem[Hermes et al.(2011)]{2011ApJ...741L..16H} Hermes, J.~J., Mullally, F., {\O}stensen, R.~H., et al.\ 2011, \apj, 741, L16
\bibitem[Hermes, et al. (2014)]{2014ApJ...789...85H} Hermes J.~J., et al., 2014, ApJ, 789, 85
\bibitem[Hermes et al.(2017)]{2017ApJS..232...23H} Hermes, J.~J., G{\"a}nsicke, B.~T., Kawaler, S.~D., et al.\ 2017, \apjs, 232, 23 
\bibitem[Hesser et al. (1976)]{1976ApJ...209..853H} Hesser J.~E., Lasker B.~M., Neupert H.~E., 1976, ApJ, 209, 853
%\bibitem[Holberg, \& Bergeron(2006)]{2006AJ....132.1221H} Holberg, J.~B., \& Bergeron, P.\ 2006, \aj, 132, 1221
\bibitem[Kepler et al. (2019)]{2019MNRAS.486.2169K} Kepler S.~O., et al., 2019, MNRAS, 486, 2169
\bibitem[Kepler \& Romero (2017)]{2017EPJWC.15201011K} Kepler S.~O., Romero A.~D., 2017, EPJWC,  01011, EPJWC.152
\bibitem[Kepler et al.(2016)]{2016MNRAS.455.3413K} Kepler, S.~O., Pelisoli, I., Koester, D., et al.\ 2016, \mnras, 455, 3413 
\bibitem[Kepler, et al. (2012)]{2012ApJ...757..177K} Kepler S.~O., et al., 2012, ApJ, 757, 177
\bibitem[Kepler, et al. (2005)]{2005A&A...442..629K} Kepler S.~O., et al., 2005, A\&A, 442, 629
\bibitem[Kepler(1993)]{1993BaltA...2..515K} Kepler, S.~O.\ 1993, Baltic Astronomy, 2, 515 
\bibitem[Kleinman et al.(2013)]{2013ApJS..204....5K} Kleinman S.~J., et al., 2013, ApJS, 204, 5
\bibitem[Koester(2010)]{2010MmSAI..81..921K} Koester, D.\ 2010, \memsai, 81, 921 
%\bibitem[Kowalski, \& Saumon(2006)]{2006ApJ...651L.137K} Kowalski, P.~M., \& Saumon, D.\ 2006, \apjl, 651, L137
\bibitem[Lenz \& Breger(2004)]{2004IAUS..224..786L} Lenz, P., \& Breger, M.\ 2004, The A-Star Puzzle, 224, 786 
\bibitem[Mason et al.(2017)]{2017AJ....154..200M} Mason, B.~D., Hartkopf, W.~I., \& Miles, K.~N.\ 2017, \aj, 154, 200
\bibitem[Montgomery et al.(2010)]{2010ApJ...716...84M} Montgomery, M.~H., Provencal, J.~L., Kanaan, A., et al.\ 2010, \apj, 716, 84
\bibitem[Montgomery et al.(2003)]{2003MNRAS.344..657M} Montgomery, M.~H., Metcalfe, T.~S., \& Winget, D.~E.\ 2003, \mnras, 344, 657
\bibitem[Mukadam et al.(2006)]{2006ApJ...640..956M} Mukadam, A.~S., Montgomery, M.~H., Winget, D.~E., Kepler, S.~O., \& Clemens, J.~C.\ 2006, \apj, 640, 956 
\bibitem[Mukadam et al. (2004)]{2004ApJ...607..982M} Mukadam A.~S., et al., 2004, ApJ, 607, 982
\bibitem[Mullally, et al. (2005)]{2005ApJ...625..966M} Mullally F., et al., 2005, ApJ, 625, 966
\bibitem[Ourique et al.(2019)]{2019MNRAS.482..649O} Ourique, G., Romero, A.~D., Kepler, S.~O., et al.\ 2019, \mnras, 482, 649
\bibitem[Renedo et al. (2010)]{2010ApJ...717..183R} Renedo I., Althaus L.~G., Miller Bertolami M.~M., Romero A.~D., C{\'o}rsico A.~H., Rohrmann R.~D., Garc{\'\i}a-Berro E., 2010, ApJ, 717, 183
\bibitem[Ricker et al.(2014)]{2014SPIE.9143E..20R} Ricker, G.~R., Winn, J.~N., Vanderspek, R., et al.\ 2014, Space Telescopes and Instrumentation 2014: Optical, Infrared, and Millimeter Wave, 914320
\bibitem[Romero et al. (2019)]{2019MNRAS.484.2711R} Romero A.~D., Kepler S.~O., Joyce S.~R.~G., Lauffer G.~R., C{\'o}rsico A.~H., 2019, MNRAS, 484, 2711
\bibitem[Romero et al. (2017)]{2017ApJ...851...60R} Romero A.~D., et al., 2017, ApJ, 851, 60
\bibitem[Romero, Campos \& Kepler (2015)]{2015MNRAS.450.3708R} Romero A.~D., Campos F., Kepler S.~O., 2015, MNRAS, 450, 3708
\bibitem[Romero et al. (2013)]{2013ApJ...779...58R} Romero A.~D., Kepler S.~O., C{\'o}rsico A.~H., Althaus L.~G., Fraga L., 2013, ApJ, 779, 58
\bibitem[Romero et al. (2012)]{2012MNRAS.420.1462R} Romero A.~D., C{\'o}rsico A.~H., Althaus L.~G., Kepler S.~O., Castanheira B.~G., Miller Bertolami M.~M., 2012, MNRAS, 420, 1462
\bibitem[Rowan et al.(2019)]{2019MNRAS.486.4574R} Rowan, D.~M., Tucker, M.~A., Shappee, B.~J., et al.\ 2019, \mnras, 486, 4574
\bibitem[Silvotti et al. (2005)]{2005AA...443..195S} Silvotti R., Voss B., Bruni I., Koester D., Reimers D., Napiwotzki R., Homeier D., 2005, A\&A, 443, 195
\bibitem[Su et al.(2017)]{2017ApJ...847...34S} Su, J., Fu, J., Lin, G., et al.\ 2017, \apj, 847, 34 
\bibitem[Taylor(2005)]{2005ASPC..347...29T} Taylor, M.~B.\ 2005, Astronomical Data Analysis Software and Systems XIV, 29
\bibitem[Tremblay et al.(2019)]{2019MNRAS.482.5222T} Tremblay, P.-E., Cukanovaite, E., Gentile Fusillo, N.~P., et al.\ 2019, \mnras, 482, 5222
\bibitem[Tremblay et al.(2013)]{2013A&A...552A..13T} Tremblay, P.-E., Ludwig, H.-G., Steffen, M., \& Freytag, B.\ 2013, \aap, 552, A13 
%\bibitem[Tremblay et al.(2011)]{2011ApJ...730..128T} Tremblay, P.-E., Bergeron, P., \& Gianninas, A.\ 2011, \apj, 730, 128
\bibitem[Voss et al. (2006)]{2006AA...450.1061V} Voss B., Koester D., Ostensen R., Kepler S.~O., Napiwotzki R., Homeier D., Reimers D., 2006, A\&A, 450, 1061
\bibitem[Winget et al. (1982)]{1982ApJ...252L..65W} Winget D.~E., van Horn H.~M., Tassoul M., Fontaine G., Hansen C.~J., Carroll B.~W., 1982, ApJ, 252, L65
\bibitem[Zhang, Robinson \& Nather (1986)]{1986ApJ...305..740Z} Zhang E.-H., Robinson E.~L., Nather R.~E., 1986, ApJ, 305, 740

\end{thebibliography}
\end{document}